\documentclass[aps,prl,twocolumn,showpacs,superscriptaddress,floatfix,10pt]{revtex4-1}
\usepackage{framed}
\usepackage{graphicx}
\usepackage{amsmath}
\usepackage{epsfig}
\usepackage{helvet}
\usepackage{amssymb}
\usepackage{framed}

\newcommand{\bra}[1]{\mbox{$\langle #1 |$}}
\newcommand{\ket}[1]{\mbox{$| #1 \rangle$}}
\newcommand{\braket}[2]{\mbox{$\langle #1 | #2 \rangle$}}

\newcommand{\proj}[1]{\mbox{$| #1 \rangle\langle #1 |$} }

\newcommand{\tr}{\mbox{$\text{tr}$}}

\def\SM{Ref. }

\begin{document}

\title{Local scale transformations on the lattice \\with tensor network renormalization} 

\author{G. Evenbly}
\affiliation{
Department of Physics and Astronomy, University of California, Irvine, CA 92697-4575 USA
}

\author{G. Vidal}
\affiliation{Perimeter Institute for Theoretical Physics, Waterloo, Ontario N2L 2Y5, Canada}  

\date{\today}

\begin{abstract}
Consider the partition function of a classical system in two spatial dimensions, or the Euclidean path integral of a quantum system in two space-time dimensions, both on a lattice. We show that the tensor network renormalization (TNR) algorithm [\emph{G. Evenbly and G. Vidal, Phys. Rev. Lett. 115, 180405}] can be used to implement local scale transformations on these objects, namely a lattice version of conformal maps. Specifically, we explain how to implement the lattice equivalent of the logarithmic conformal map that transforms the Euclidean plane into a cylinder. As an application, and with the 2D critical Ising model as a concrete example, we use this map to build a lattice version of the scaling operators of the underlying conformal field theory, from which one can extract their scaling dimensions and operator product expansion coefficients.
\end{abstract}

\pacs{05.30.-d, 02.70.-c, 03.67.Mn, 75.10.Jm}

\maketitle

Wilson's renormalization group (RG) \cite{Wilson}, key to our understanding of statistical mechanics and quantum field theory, has the notion of scale transformation at its very core. Under RG or scale transformations, a generic system flows in the space of theories until it reaches a fixed point. The different fixed points of this RG flow, each corresponding to a scale invariant theory, play a particularly important role: they enumerate the possible phases and phase transitions of the system, while encapsulating their universal properties.
On the other hand, a local field theory that is invariant under \textit{global} changes of scale is often also invariant under \textit{local} scale transformations --- and, more generally, the whole conformal group \cite{Polchinski}. Accordingly,
due to its connection to critical phenomena (but also its role in string theory), local scale invariance and conformal field theory (CFT) \cite{CFT} have been intensively studied over the years.

RG methods also have a long history on the lattice, with Kadanoff's spin-blocking procedure \cite{Kadanoff} pre-dating Wilson conceptualization of the RG flow. Substantial progress in real-space methods has been possible during the last decade by reformulating the RG program using tensor networks \cite{TRG, MERA, TRGenv, SRG, TEFR, HOTRG, TNR, TNRtoMERA, MPS}. On the algorithmic side, Levin and Nave's ground-breaking \textit{tensor renormalization group} (TRG) algorithm \cite{TRG} made it possible to numerically coarse-grain the statistical partition function of a \textit{generic} 2D classical system (as well as the Euclidean path integral of a quantum system in 1+1 dimensions). On the conceptual side, however, a scale or RG transformation on the lattice should be required to be consistent with the known structure of the RG flow in the continuum. The \textit{tensor network renormalization} (TNR) algorithm \cite{TNR} was recently developed for this purpose. Thanks to the use of disentanglers that properly dispose of irrelevant, short-ranged degrees of freedom during the coarse-graining transformation, TNR implements a global scale transformation that reproduces the expected structure of the RG flow and its fixed points \cite{fpTEFR}. In particular, it explicitly recovers scale invariance at criticality, in the form of a critical fixed-point tensor.

The goal of this paper is to show that TNR can also properly implement \textit{local} scale transformations, and thus study local scale invariance directly on the lattice. Here, local scale invariance is to be understood as an approximate, emergent symmetry of a lattice system at criticality. For concreteness, we will mostly explain how to perform the lattice version of a specific local scale transformation that maps the plane into a cylinder. Moving along the axis of this cylinder corresponds to changing scale in the plane. We numerically demonstrate with the 2D classical Ising model that, at a continuous phase transition, two well-known facts in the continuum also hold on the lattice: (i) \textit{scale invariance} on the plane results in \textit{translation invariance} on the cylinder; (ii) from the spectrum of eigenvalues of a transfer matrix $\mathcal{R}$ on the cylinder, an accurate numerical estimate of the scaling dimensions $\Delta_{\alpha}$ of the underlying CFT can be extracted. It is a highly non-trivial consequence of local scale invariance that the eigenvalues of $\mathcal{R}$ are closely related to those of another transfer matrix $\mathcal{M}$ obtained by directly placing the original lattice model on a cylinder \cite{Cardy,CFT}. Therefore, the demonstrated ability to accurately reproduce the relation between the spectra of $\mathcal{R}$ and $\mathcal{M}$ is strong evidence that TNR has properly implemented the local scale transformation. Our construction also yields an explicit lattice representation of the scaling operators of the underlying CFT, from which we can then extract other universal data, such as the central charge or the operator product expansion (OPE) coefficients \cite{CFT}.
 
\begin{figure}[!t]
\begin{center}
\includegraphics[width=7cm]{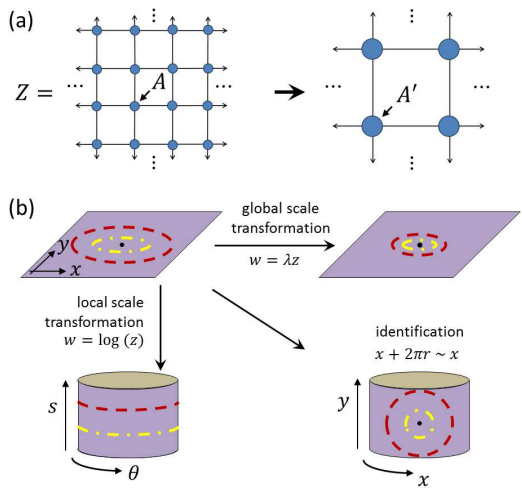}
\caption{ 
(a) Graphical illustration of the tensor network, made of copies of a tensor $A$, that represents the partition function $Z$ of a lattice model. Tensor network renormalization (TNR) produces a coarser tensor network made of tensors $A'$.
(b) The plane $z = x+iy$ is mapped into itself by a global scale transformation $w = \lambda z$ with constant re-scaling factor $\lambda$, or to an infinite cylinder of unit radius by a local scaling transformation $w = \log(z)$ with local re-scaling factor $\lambda = 1/|z|$. In this second case, a circle $|z|^2 =  x^2 + y^2$ on the plane is mapped into a circle at constant $s = \log_2 |z|$ and $\theta \in [0,2\pi)$ on the cylinder, where $w = s + i \theta$, and a translation $s \rightarrow s+a$ on the cylinder can be interpreted as a change of scale $z \rightarrow e^{a}z$ in the plane.
Alternatively, an infinite cylinder of finite radius $r$ can also be obtained from the plane $z=x+iy$ by e.g. identifying the coordinate $x$ with $x+2\pi r$.
}
\label{fig:localscale}
\end{center}
\end{figure}

\textit{Change of scale on the lattice.---} For concreteness, let us consider the partition function $Z$ of a translation invariant system on the square lattice. This partition function can be canonically represented \cite{TRG, TRGenv, SRG, TEFR, HOTRG, TNR} as a tensor network made of copies of a single tensor $A$, see Fig. \ref{fig:localscale}(a). [An example of $Z$ and $A$ can be found in Eqs. \ref{eq:Ising}-\ref{eq:AIsing}.] Tensor network renormalization (TNR) \cite{TNR} can then be used to perform a global (discrete) scale transformation, 
\begin{equation} \label{eq:global}
(x,y) \rightarrow (x',y') = (\lambda x, \lambda y),
\end{equation}
where $x$ and $y$ label the lattice sites and $\lambda = 1/2$ is the re-scaling factor. This is accomplished by acting on the tensor network with a sequence of local manipulations that effectively replace blocks of $2\times 2$ tensors $A$ with new coarse-grained tensors $A'$, see Fig. \ref{fig:localscale}. The intricate details of these manipulations can be found in Refs. \cite{TNR,TNRalgorithms} and have been briefly summarized in \SM \cite{SM} Section A for completeness. However, for our current purposes we only need to focus on one salient feature of these manipulations. Namely, the global change of scale is implemented through a sequence of \textit{local} replacements, so that the coarse-graining can in principle be applied in one region of the network while leaving the rest unchanged. Accordingly, we should also be able to perform \textit{local} scale transformations, with a scaling factor $\lambda$ in Eq. \ref{eq:global} that now depends on the coordinates $(x,y)$.


\begin{figure}[!t]
\begin{center}
\includegraphics[width=8cm]{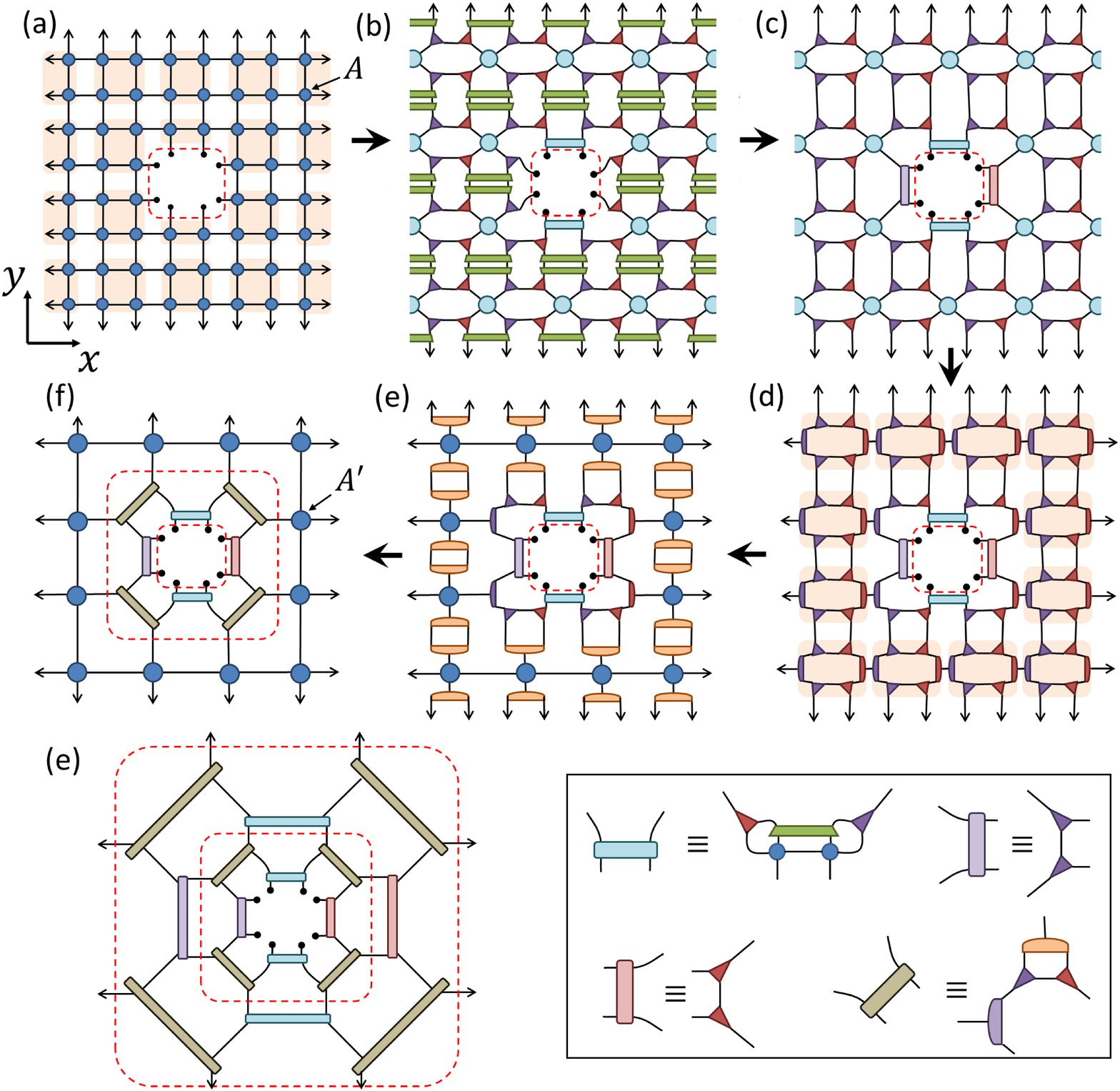}
\caption{
(a) Tensor network obtained by removing four tensors $A$ from the representation of the partition function $Z$ on the plane shown in Fig. \ref{fig:localscale}(a). The puncture at the origin of the plane leaves eight open indices.
(b)-(f) Applying TNR to implement a global scale transformation on the plane (see \SM \cite{SM} Section A) but without touching the open indices results in a new square tensor network with a puncture, made of tensors $A'$, together with two concentric rows of tensors around the origin. The inset relates these tensors to the auxiliary tensors that appear while coarse-graining the bulk.
}
\label{fig:TNRlocal1}
\end{center}
\end{figure}

\textit{Logarithmic map.---} 
Specifically, here we will study a lattice version of the logarithmic conformal map employed in CFT in the context of radial quantization \cite{CFT} and illustrated in Fig. \ref{fig:localscale}. As any conformal map, it can be decomposed as a change of coordinates followed by a local re-scaling. Consider first the change of coordinates $(x,y) \rightarrow (s,\theta)$, 
\begin{equation}\label{eq:log}
s \equiv \log(\sqrt{x^2 + y^2}),~~~~\theta \equiv \arctan \frac{y}{x},
\end{equation}
where $s \in \mathbb{R}$ represents length scale and $\theta$ is an angle, $\theta \in [0,2\pi)$. This change of variables can be compactly written as a logarithm, $z \rightarrow w = \log (z)$, using complex coordinates $z \equiv x + i y$ and $w \equiv s + i \theta$. Coordinates $(s,\theta)$ parametrize a cylinder, but with a non-uniform metric, since infinitesimal distances read $dx^2 + dy^2 = e^{2s} \left(d\theta^2 + ds^2\right)$. We then eliminate the factor $e^{2s}$ by means of a local change of scale (Weyl re-scaling) with re-scaling factor 
\begin{equation} \label{eq:Weyl}
\lambda_{(x,y)} = \frac{1}{\sqrt{x^2+y^2}} = e^{-s},
\end{equation}
so that $e^{2s} \left(d\theta^2 + ds^2\right) \rightarrow d\theta^2 + ds^2$.
Thus, the change of variables and local re-scaling transformation in Eqs. \ref{eq:log}-\ref{eq:Weyl} map the plane $(x,y)$ into a cylinder $(s,\theta)$ of unit radius. 

\textit{From the plane to the cylinder, on the lattice.---} Let us now explain how the above logarithmic transformation can be implemented on the lattice partition function $Z$, see Fig. \ref{fig:TNRlocal1}. First, from the initial tensor network representation of $Z$ we remove a block of $2\times 2$ tensors $A$ encircling the origin $(x,y) = (0,0)$. We think of the resulting network, which contains a hole with eight open indices, as representing the punctured plane. Then we apply the homogeneous coarse-graining transformation of Ref. \cite{TNR} (see \SM \cite{SM} Section A) everywhere in the network except near the puncture, where the eight open indices are not touched. The net result is a uniform change of scale by a factor $\lambda = 1/2$ everywhere except in a neighbourhood of the origin, where no change of scale has been performed, that is where $\lambda = 1$. 

\begin{figure}[!t]
\begin{center}
\includegraphics[width=7.5cm]{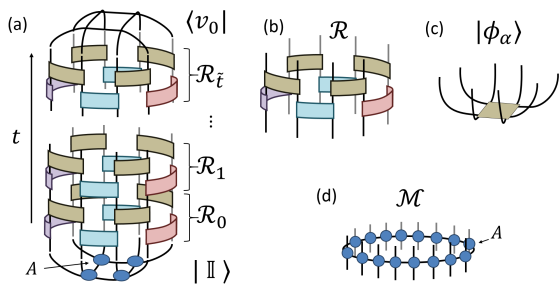}
\caption{
(a) Partition function $Z$ expressed in terms of the resulting cylindrical tensor network, Eq. \ref{eq:Zcyl}. At the basis of the cylinder, a block of four tensors $A$, denoted $\ket{\mathbb{I}}$, has been re-inserted to remove the puncture previously introduced at origin of the plane. On a finite system, the cylinder is also finite and has a set of lines, denoted $\bra{v_0}$, at the top.
(b) Critical scaling map $\mathcal{R}$ in Eq. \ref{eq:R}. 
(c) Eigenvector $\ket{\phi_{\alpha}}$ of the scaling map $\mathcal{R}$.
(d) Transfer matrix $\mathcal{M}$ in Eq. \ref{eq:M}.
}
\label{fig:TNRlocal2}
\end{center}
\end{figure}

The resulting tensor network, depicted in Fig. \ref{fig:TNRlocal1}(f), is made of two parts: (i) two concentric rows of tensors around the puncture at the origin and (ii) a square tensor network made of coarse-grained tensors $A'$ throughout the plane, but with four tensors $A'$ missing at the origin. Notice that this second network is just a re-scaled version of the punctured tensor network we started from, Fig. \ref{fig:TNRlocal1}(a). Therefore we can again apply the coarse-graining transformation of Fig. \ref{fig:TNRlocal1}(b)-(f) to it. This results in an overall re-scaling by a factor $\lambda = (1/2)^2$ away from the puncture, Fig. \ref{fig:TNRlocal1}(e). Iterating this procedure further, we obtained a coarse-grained tensor network in which regions of the square lattice at a distance $ \sqrt{x^2+y^2} \approx 2^t$ from the origin have indeed been re-scaled by an overall factor $\lambda = (1/2)^{t} = 1/\sqrt{x^2 + y^2}$ as in Eq. \ref{eq:Weyl}.
[Notice that, due to the discrete nature of the possible scaling factors on the lattice, it is more convenient to parametrize scale by an integer $t \equiv \log_2(|z|)$ which relates to the continuous scale parameter $s = \log(|z|)$ in Eq. \ref{eq:log} by $2^t = e^s$, or $s = t \log(2)$.]

\textit{Discretized cylinder and transfer matrix.---} 
The initial partition function $Z$ can be re-written as the product
\begin{equation} \label{eq:Zcyl}
Z = \bra{v_{0}} \mathcal{R}_{\tilde{t}} \cdots \mathcal{R}_1 \mathcal{R}_0 \ket{\mathbb{I}}, 
\end{equation}
where, as shown in Fig. \ref{fig:TNRlocal2}(a), $\mathcal{R}_t$ denotes a linear map made of two rows of tensors that implements a change of scale in the plane, from length $2^{t}$ to $2^{t+1}$. For instance, one can check that, by construction, acting with $\mathcal{R}_t$ on a block of $2\times 2$ tensors $A^{(t)}$, denoted $\ket{\mathbb{I}}$, produces a block of $2\times 2$ tensors $A^{(t+1)}$ (see \SM \cite{SM} Section A). The vector $\bra{v_{0}}$ is used to perform a trace on the eight open indices of a block of $2\times 2$ tensors. The height of the cylinder is given by $\tilde{t} \approx \log_2 N$ if the initial tensor network was made of $N\times N$ tensors $A$ (with periodic boundary conditions), and is thus infinite in the thermodynamic limit. 

\begin{figure}[!t]
\begin{center}
\includegraphics[width=6cm]{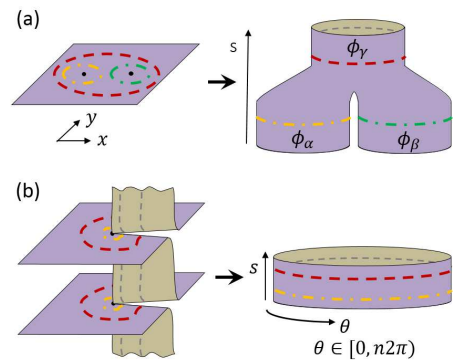}
\caption{
(a) A local scale transformation maps the plane with two punctures into a pair of pants. On the lattice, from this pair of pants we can extract the OPE coefficient $C_{\alpha \beta \gamma}$ by inserting scaling operators $\ket{\phi_{\alpha}}$ and $\ket{\phi_{\beta}}$ on the punctures and evaluating the amplitude of obtaining scaling operator $\ket{\phi_{\gamma}}$ as a result of their fusion.
(b) A logarithmic conformal transformation maps an $n$-sheeted Riemann surface, consisting of $n$ copies (here, $n=2$) of the plane periodically connected through semi-infinite branch cuts along the positive $x$ axis to a cylinder ($s,\theta$) with $s\in \mathbb{R}$ and $\theta \in [0, n2\pi)$.
}
\label{fig:otherLST}
\end{center}
\end{figure}

At a critical point, where TNR explicitly realizes scale invariance \cite{TNR}, the above linear maps $\mathcal{R}_{t}$ is seen to become independent of the scale $t$ \cite{R} and, accordingly, the tensor network on the cylinder has become translation invariant. Let $\mathcal{R}$ denote the critical transfer matrix on the cylinder, see Fig. \ref{fig:TNRlocal2}(b). From its eigenvectors $\ket{\phi_{\alpha}}$,
\begin{equation} \label{eq:R}
 \mathcal{R} \ket{\phi_{\alpha}} = r_{\alpha} \ket{\phi_{\alpha}},~~~~~r_{\alpha} \equiv 2^{-\Delta_{\alpha}},
\end{equation} 
we then obtain an explicit lattice representation, in the form of a tensor with eight indices depicted in Fig. \ref{fig:TNRlocal2}(c), of the scaling operators $\phi_{\alpha}$ of the theory, while the eigenvalues $r_{\alpha}$ yield a numerical estimate of their scaling dimensions $\Delta_{\alpha}$. 

\textit{Other geometries.---} From the lattice representation $\ket{\phi_{\alpha}}$ of the primary fields we can also compute the OPE coefficients $C_{\alpha\beta\gamma}$ of the underlying CFT. This is accomplished by considering another local coarse-graining transformation, depicted in Fig. \ref{fig:otherLST}(a), that produces a pair of pants geometry describing the fusion of two scaling fields into a third one (see \SM \cite{SM} Section C). On the other hand, the central charge $c$ of the CFT can be obtained by applying the logarithmic map to a lattice version of an $n$-sheeted Riemann surface, depicted in Fig. \ref{fig:otherLST}(b), which produces a thicker cylinder (see \SM \cite{SM} Section D).

\begin{figure}[!t]
\begin{center}
\includegraphics[width=8cm]{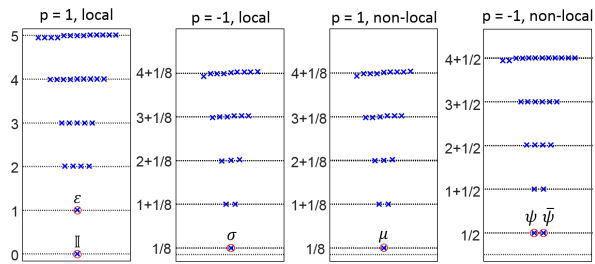}
\caption{
First 101 scaling dimensions $\Delta_{\alpha}$ of the critical Ising model, obtained from the eigenvalues $2^{-\Delta_{\alpha}}$ of the scaling maps $\mathcal{R}$ (first and second columns) and $\mathcal{R}'$ (third and fourth columns, see \SM \cite{SM} Section B). Even and odd spin $\mathbb{Z}_2$ symmetry sectors are indicated by $p = +1$ and $p=-1$, respectively. The primary fields identity $\mathbb{I}$, spin $\sigma$, and energy density $\epsilon$, together with their descendants, are the local scaling operators; the disorder $\mu$ and the fermions $\psi$ and $\bar{\psi}$ and their descendants are non-local scaling operators. 
}
\label{fig:IsingDims}
\end{center}
\end{figure}

\textit{Example.---} Let us consider the critical 2D Ising model 
\begin{equation} \label{eq:Ising}
Z = \sum_{\{\sigma\}} e^{-H(\{\sigma\})/T_c}, ~~~H\left( \{ \sigma \} \right) =  - \sum_{\left\langle i,j \right\rangle}  \sigma _i\sigma _j
\end{equation}
on the square lattice, where $\sigma \in \{+1,-1\}$ is an Ising spin and $T_c \equiv 2/\log(1+\sqrt{2})$. In components, tensor $A$ reads 
\begin{equation} \label{eq:AIsing}
A_{ijkl} = \exp\left[(\sigma_i\sigma_j + \sigma_j\sigma_k + \sigma_k\sigma_l + \sigma_l\sigma_i)/T_c\right].
\end{equation}
We first use TNR to apply a global scale transformation with scaling factor $\lambda = 1/2$ several times as described in Ref. \cite{TNR, TNRalgorithms}, realizing scale invariance. With bond dimension $\chi=6$, this takes less than one minute on a dual core laptop. Then, we use the auxiliary, scale-invariant tensors in the insets of Fig. \ref{fig:TNRlocal1}, to build the transfer matrix $\mathcal{R}$ on the cylinder. Fig. \ref{fig:IsingDims} shows the scaling dimensions $\Delta_{\alpha}$ of local scaling operators obtained from $\mathcal{R}$. For instance, for the local primary fields spin $\sigma$ and energy density $\epsilon$ we obtain $\Delta_{\sigma} \approx 0.124$, and $\Delta_{\epsilon} = 1.001$, very close to the exact values $0.125$ and $1$ (the exact $\Delta_{\mathbb{I}}=0$ results trivially from the normalization of tensor $A$). Using the geometries in Fig. \ref{fig:otherLST} we also obtain the non-vanishing OPE coefficient between these primaries, namely $C_{\sigma\sigma\epsilon}\approx0.5001$, and central charge $c\approx 0.5003$, again close to their exact value $0.5$.

\textit{Discussion.---} Let us consider now the transfer matrix $\mathcal{M}$ of Fig. \ref{fig:TNRlocal2}(d), which is built by connecting together $N$ of the initial tensors $A$ into a ring. At criticality, the spectrum of eigenvalues of $\mathcal{M}$ is also given in terms of the scaling dimensions $\Delta_{\alpha}$ of the theory \cite{Cardy, CFT, TEFR, M},
\begin{equation} \label{eq:M}
 \mathcal{M} \ket{\psi_{\alpha}} = m_{\alpha} \ket{\psi_{\alpha}}, ~~~m_{\alpha} \equiv e^{-\frac{2\pi}{N} (\Delta_{\alpha} - \frac{c}{12})}.
\end{equation} 
The detailed, one-to-one relation between the eigenvalues of the transfer matrix $\mathcal{M}$, made of tensors $A$, and of the transfer matrix $\mathcal{R}$, which was obtained after applying a local scale transformation to the network made of tensors $A$, is a highly non-trivial manifestation of local scale/conformal invariance in the continuum \cite{CFT}, whose derivation is reviewed in \SM \cite{SM} Section E. We regard TNR's ability to accurately reproduce this relation on the lattice as strong evidence that it has properly implemented the local scale transformation. 

We emphasize that other coarse-graining transformations, such as tensor renormalization group (TRG) and generalizations \cite{TRG, TRGenv, SRG, TEFR, HOTRG}, can also be applied locally. However, most of these methods do not thoroughly remove short-range entanglement, failing to explicitly recover global scale invariance at criticality \cite{TNR}. In particular, the logarithmic local scale transformation would produce a linear map $\mathcal{R}_t$ that depends on the scale $t$ and from which we can not extract the scaling operators and their scaling dimensions. We thus conclude that the ability of TNR to correctly implement local scale transformations can be traced back to its proper removal of short-range entanglement. 

In this paper we have mostly focused on a local scale transformation of the plane that leaves a puncture at the origin untouched, Fig. \ref{fig:TNRlocal1}. We have also briefly considered a local scale transformation that leaves two punctures untouched, Fig. \ref{fig:otherLST}(a). One can of course consider many other local scale transformations on the lattice. In particular, by leaving the whole $x$ axis of the plane untouched, we can map the upper half plane into the hyperbolic plane. This was done in Ref. \cite{TNRtoMERA}, where a tensor network representation of the statistical partition function $Z$ of a classical system in two space dimensions was converted into a multi-scale entanglement renormalization ansatz (MERA) \cite{MERA} representation of the ground state of the dual quantum theory in two space-time dimensions. 

We thank Davide Gaiotto and Rob Myers for insightful comments and discussions, and for clarifying how the normalization of the tensors representing a partition function should be adjusted under conformal transformations. G. E. and G. V. acknowledge support by the Simons Foundation (Many Electron Collaboration). G. V. acknowledges support by the John Templeton Foundation and thanks the Australian Research Council Centre of Excellence for Engineered Quantum Systems. Research at Perimeter Institute is supported by the Government of Canada through Industry Canada and by the Province of Ontario through the Ministry of Research and Innovation.

\newpage
\title{Supplementary Material: Local scale transformations on the lattice with tensor network renormalization} 

\author{G. Evenbly}
\affiliation{
Department of Physics and Astronomy, University of California, Irvine, CA 92697-4575 USA
}

\author{G. Vidal}
\affiliation{Perimeter Institute for Theoretical Physics, Waterloo, Ontario N2L 2Y5, Canada}  

\maketitle

\renewcommand\thefigure{A.\arabic{figure}}    
\renewcommand{\theequation}{A.\arabic{equation}}
\setcounter{figure}{0}  
\setcounter{equation}{0} 

\section{Section A: Scale transformations on the lattice}
In this section we review how, and in which sense, the tensor network renormalization  \cite{TNR,TNRalgorithms} procedure implements global and local scale transformations on the lattice. 

\subsection{Global scale transformation.}

\begin{figure}[!t]
\begin{center}
\includegraphics[width=8.5cm]{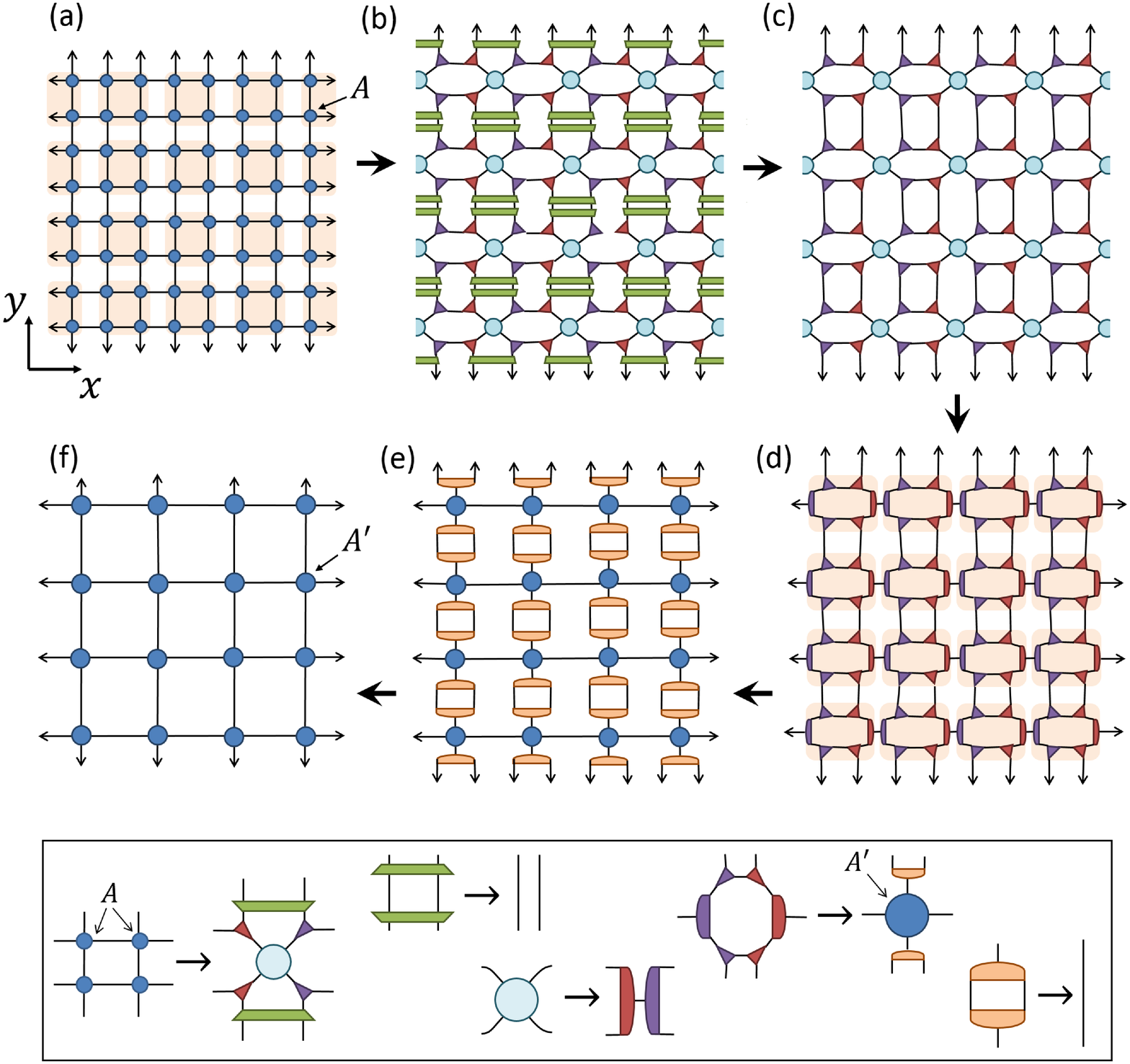}
\caption{
(a) Tensor network representing the partition function $Z$ of a statistical system on an infinite 2D lattice with coordinates ($x,y$) (equivalently, the Euclidean path integral of a quantum system in $1+1$ space-time dimensions, with discretized Euclidean time $\tau=y$). 
(b)-(f) Sequence of manipulations that implement a global change of scale. The resulting tensor network is made of copies of coarse-grained tensor $A'$, and there is one tensor $A'$ in the final tensor network for every four tensors $A$ in the initial tensor network. The inset shows the set of five local replacements required to implement such global scale transformation. 
}
\label{fig:TNRglobal}
\end{center}
\end{figure}

TNR was proposed in Ref. \cite{TNR} as a coarse-graining transformation for tensor networks. Fig. \ref{fig:TNRglobal} shows the sequence of local manipulations of the initial tensor network, made of copies of a tensor $A$, that produces a coarser tensor network, made of copies of a tensor $A'$. We refer to Refs. \cite{TNR,TNRalgorithms} for a detailed explanation of the auxiliary tensors described in Fig. \ref{fig:TNRglobal}, as well as the optimization procedure employed to determine these tensors.

The initial tensor network may be used to represent the Euclidean path integral of a quantum system in one spatial dimension or the statistical partition function of a classical system in two spatial dimensions. In this case, tensor $A$ encodes either the discretized imaginary time evolution of the quantum system or the statistical Boltzmann weights of the classical system. A regular lattice is defined by a discrete subgroup of translations, so that the initial tensor network on the square lattice is naturally endowed with a notion of discrete translations. In addition, since the network is made of copies of a single tensor $A$, it is invariant under these discrete translations. In contrast, a lattice is not automatically endowed with a notion of scale transformation, let alone of scale invariance. So, how should we go about defining such notions?

\begin{figure}[!t]
\begin{center}
\includegraphics[width=8.5cm]{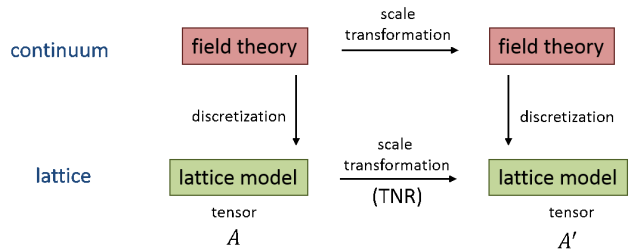}
\caption{
Tensor network renormalization can be used to define a notion of scale transformation (by a discrete scaling factor) on the lattice, which mimics the effect of a scale transformation in the continuum. When acting on a tensor network made of tensors $A$, the scale transformation produces a new network made of tensors $A'$. At a critical point, if the lattice model is described in the continuum (e.g. in the limit of large distances compared with the lattice spacing $a$) by a field theory that is invariant under changes of scale, then TNR explicitly reproduces scale invariance on the lattice, in the sense that the tensor $A$ before the coarse-graining transformation and the coarse-grained tensor $A'$ are the same, $A=A'$, up to small truncation errors that can be systematically reduced by increasing the bond dimension $\chi$.
}
\label{fig:diagram}
\end{center}
\end{figure}

From a geometric perspective, it seems natural to resort to a coarse-graining transformation in order to define a scale transformation on the lattice and/or tensor network. Indeed, replacing a block of tensors $A$ with a coarse-grained tensor $A'$ has the net effect of producing a new tensor network in which the original lattice spacing $a$ between nearest neighbour tensors $A$ has been replaced with a new lattice spacing $a'$ between nearest neighbour tensors $A'$, where $a'>a$. For instance, TNR in Fig. \ref{fig:TNRglobal} is such that $a' = 2a$.

Here we take the point of view that a proper implementation on the lattice of the notion of a scale transformation by means of a coarse-graining transformation should also satisfy some additional consistency conditions. Namely, when the lattice model can be understood as a discretized version of a field theory in the continuum, where the notion of scale transformation is well-defined, then we should require that the process of discretization of the field theory into a lattice model and the scale transformation commute, see Fig. \ref{fig:diagram}. That is, we require that our coarse-graining transformation has an effect, on the lattice, equivalent to that of a scale transformation in the continuum. 

This implies, in particular, that when the continuum field theory is invariant under global scale transformations, as in the case of a conformal field theory (CFT) \cite{CFT}, then the corresponding lattice model should also be explicitly invariant under the available (that is, discrete) global changes of scale. At the level of the tensor network, this means that when studying a critical lattice system with an underlying CFT, the initial tensor $A$ and the coarse-grained tensor $A'$ should be the same..

\begin{figure}[!t]
\begin{center}
\includegraphics[width=8.5cm]{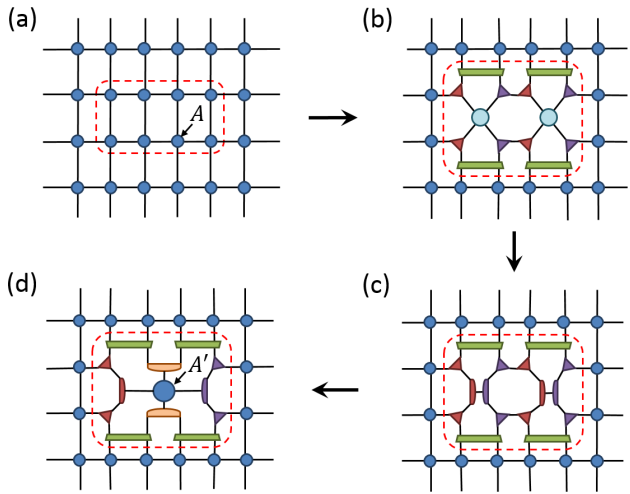}
\caption{
Due to its local character, TNR can implement a local scale transformation on a small neighborhood of the tensor network without affecting the rest of the network. 
(a) Initial tensor network with a block of $2 \times 4$ tensors $A$ to be coarse-grained.
(b) - (d) Sequence of manipulations restricted to part of the network that produce a new tensor $A'$ and several auxiliary tensors that connect $A'$ with the part of the network that has not been modified.
}
\label{fig:LocalCoarseGraining}
\end{center}
\end{figure}

In Ref. \cite{TNR}, it was numerically demonstrated for the particular case of the Ising model that, indeed, TNR explicitly realizes scale invariance at criticality, in that the tensor $A$ becomes invariant under a global scale transformation. That is, under a TNR, the coarse-graining $A \rightarrow A'$ is such that $A'=A$, up to small truncation errors that can be systematically decreased by increasing the refinement parameter $\chi$ of the numerical scheme, which corresponds to the bond dimension of the tensors. This is a strong indication that TNR can be used to define and implement a good notion of global scale transformation on the lattice, that is, a scale transformation by a rescale factor that is the same throughout the lattice. More generally, Ref. \cite{TNR} demonstrated, for the particular case of the Ising model, that TNR generates an RG flow in the space of tensors with the expected structure of fixed points, namely a critical fixed-point at the critical temperature $T_c$ and two non-critical fixed-points for temperature $T \not = T_c$  (a symmetry breaking fixed-point for $T < T_c$, and a disordered fixed-point for $T>T_c$). We emphasize that for the two non-critical RG fixed points, the fixed-point tensors that TNR obtained where the same than those previously obtained using another coarse-graining transformation, namely tensor entanglement-filtering renormalization (TEFR) in Ref. \cite{TEFR}.

\subsection{Local scale transformation.} 
 
The purpose of this paper is to investigate whether TNR also allows us to define a proper notion of local scale transformation on the lattice. Since TNR consists of local manipulations of the tensor network, it allow us to implement a change of scale in on region while leaving the rest of the network untouched. This is illustrated in Fig. \ref{fig:LocalCoarseGraining}, where the manipulation of an initial square lattice of tensors $A$ only affects a group of eight such tensors. The net result is a single tensor $A'$ connected to the rest of the network made of tensors $a$ by means of a number of auxiliary tensors. 

What condition should we require on TNR to be able to state that this local coarse-graining of the tensor network has properly implemented a local scale transformation? A natural requirement would be that the commutativity expressed in the diagram of Fig. \ref{fig:diagram}(a) holds also for local scale transformations. Then in particular, a necessary condition would be that, when used to mimic a local scale transformation (or conformal transformation) in a continuous theory that is invariant under local scale transformations (such as a conformal field theory), TNR should produce a new tensor network that, again, corresponds to a discretized version of the transformed field theory, so that the diagram of Fig. \ref{fig:diagram}(b) holds also for local scale transformations. 

\begin{figure}[!t]
\begin{center}
\includegraphics[width=8cm]{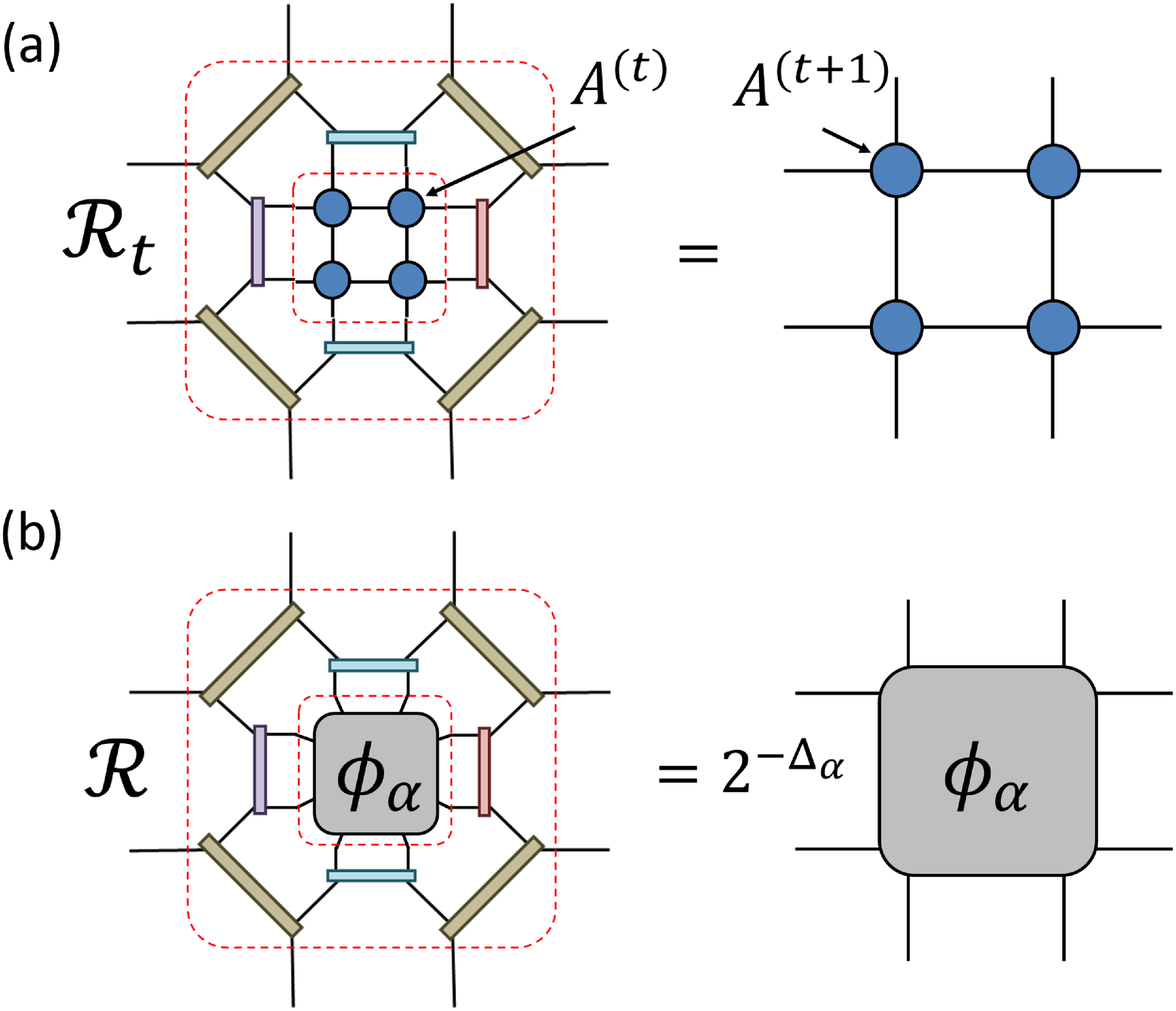}
\caption{
(a) The linear map $\mathcal{R}_t$, when applied to a $2 \times 2$ block of tensors $A^{(t)}$, produces a block of $2 \times 2$ tensors $A^{(t+1)}$. Therefore $\mathcal{R}_t$ implements a translation along the cylinder corresponding to a change of scale in the plane.
(b) At criticality, $\mathcal{R}_t$ no longer depends on the scale $t$ and becomes the scaling map $\mathcal{R}$. The eigenvectors $\ket{\phi_{\alpha}}$ of $\mathcal{R}$ and eigenvalues $r_{\alpha} = 2^{-\Delta_{\alpha}}$ correspond to the scaling operators $\phi_{\alpha}$ and scaling dimensions $\Delta_{\alpha}$ of the underlying CFT.}
\label{fig:ScalingOperators}
\end{center}
\end{figure}

In this paper we establish that this is indeed the case for a particular local scale transformation, namely the conformal map that takes the plane into a cylinder. As explained in the main text, in a non-critical system the resulting cylinder is described as the product of linear maps $\mathcal{R}_0$, $\mathcal{R}_1$, $\cdots$. Each of these maps implements a translation on the cylinder, which corresponds to changing scale on the plane, see Fig. \ref{fig:ScalingOperators}(a). At criticality, scale invariance on the plane implies translation invariance on the cylinder, and the spectrum of the resulting transfer matrix $\mathcal{R}$ is organized according to the same scaling dimensions $\Delta_{\alpha}$ of the continuous CFT, see Fig. \ref{fig:ScalingOperators}(b). Moreover, the eigenvectors $\ket{\phi_{\alpha}}$ of $\mathcal{R}$ provide a lattice realization of the scaling operators $\phi_{\alpha}$ of the CFT, as witnessed by the fact that their fusion at short distances occurs according to the correct operator product expansion (OPE) coefficients of the CFT (see Section C).

\renewcommand\thefigure{B.\arabic{figure}}    
\renewcommand{\theequation}{B.\arabic{equation}}
\setcounter{figure}{0}  
\setcounter{equation}{0} 

\section{Section B: Non-local scaling operators.} 

\begin{figure}[!t]
\begin{center}
\includegraphics[width=8cm]{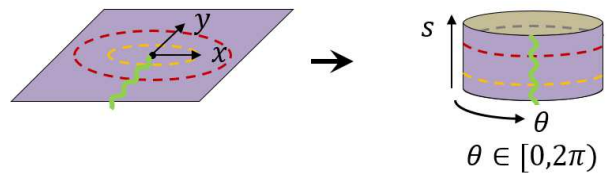}
\caption{
When applied the plane with a semi-infinite defect line starting at the origin, the logarithmic transformation produces a cylinder with a defect line in the $s$ direction.
}
\label{fig:nonlocal}
\end{center}
\end{figure}

\begin{figure}[!t]
\begin{center}
\includegraphics[width=8.5cm]{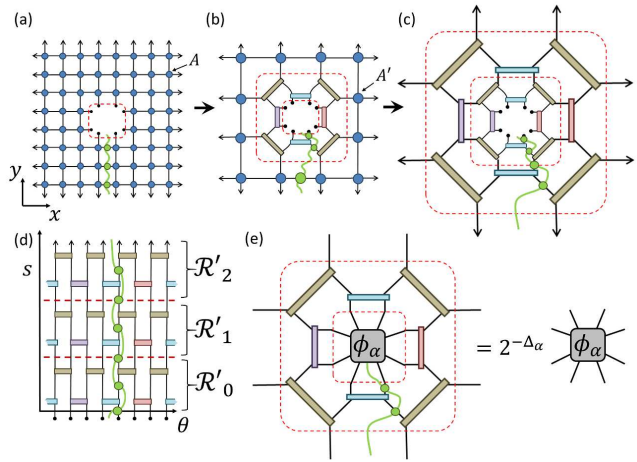}
\caption{
(a) Tensor network representation of the partition function of the 2D Ising model on the plane with a semi-infinite line defect from the origin (that follows the negative half of the $y$-axis), and where we have added a puncture (removed four tensors) at the origin of the plane. The semi-infinite line defect corresponds to anti-periodic boundary conditions and is implemented by acting on the appropriate indices with a matrix that implements the $\mathbb{Z}_2$ spin-flip symmetry of the model.
(b) The result of applying a homogeneous coarse-graining transformation throughout the tensor network except at the puncture is analogous to that in the absence of the line defect, except for the presence of the matrices implementing the $\mathbb{Z}_2$ symmetry. Using $\mathbb{Z}_2$ symmetric tensors, which commute with those matrices, one can see that the line defect survives the coarse-graining. 
(c) Iteration of the coarse-graining transformation produces a cylinder with a line defect.
(d) The cylinder can be understood as the product of linear maps $\mathcal{R}'_0$, $\mathcal{R}'_1$, $\mathcal{R}'_2$, etc. Each linear map $\mathcal{R}'_{t}$  is made of the same tensors as $\mathcal{R}_{t}$ in the absence of the line defect, but with an addition pair of matrices that implement anti-periodic boundary conditions.
(e) At criticality, from the scaling map $\mathcal{R}'$ we can extract a lattice representation of non-local scaling operators $\phi_{\alpha}$ and their scaling dimension $\Delta_{\alpha}$.
}
\label{fig:TNRnonlocal}
\end{center}
\end{figure}

In this section we apply the logarithmic local scale transformation discussed in the main text to the partition function of the Ising model with a line defect that goes from the origin to infinite, see Figs. \ref{fig:nonlocal}-\ref{fig:TNRnonlocal}. 

Specifically, we consider the case where this line defect implements antiperiodic boundary conditions, that is, across the defect line the sign of the nearest neighbor Hamiltonian term $-\sigma_i\sigma_j$ is changed to $+ \sigma_i \sigma_j$. More general line defects can be similarly addressed, including the Kramers-Wannier duality defect analysed in Ref. \cite{sHauru}. 

Fig. \ref{fig:TNRnonlocal}(a) shows a tensor network representation of the partition function on the plane with a puncture at the origin and antiperiodic boundary conditions running from the puncture downwards along the semi-infinite line $x=0$ and $y<0$. The antiperiodic boundary conditions can be implemented by inserting a matrix, corresponding to a spin flip (the same that implements the global $\mathbb{Z}_2$ symmetry of the model) on each index crossing the negative half of the $y$-axis. 

The tensor network described above, with a puncture and a semi-infinite line defect, can be coarse-grained in a very similar way as we did in the absence of the line defect. In particular, away from the defect we can use the same tensor replacements as before. At the defect, if $\mathbb{Z}_2$ symmetric tensors are used, careful consideration of their symmetry properties shows that we can again use the same tensor replacements as before, only modified by the introduction of matrices that implement the $\mathbb{Z}_2$ symmetry, in the same spirit of what occurs when considering non-local operators with the MERA \cite{sMERAnonlocal}. The result is a map to a discretized cylinder, see Fig. \ref{fig:TNRnonlocal}(d), which inherits the anti-periodic boundary conditions. This cylinder can be written as the product of linear maps $\mathcal{R}'_0$, $\mathcal{R}'_1$, $\cdots$, where the map $\mathcal{R}'_{t}$ is made of the same tensors as the map $\mathcal{R}_{t}$ in the main text, except for the anti-periodic boundary conditions. The linear map $\mathcal{R}'_{t}$ implements a change of scale on operators that consist of a local core (placed at the origin of the plane) followed by a semi-infinite line defect. 

At criticality, we observe again that the cylinder becomes translation invariant --that is, $\mathcal{R}'_{t}$ does not depend on $s$. Let $\mathcal{R}'$ denote this new (non-local) scaling map. Diagonalizing $\mathcal{R}'$ we obtain a lattice representation $\ket{\phi_{\alpha}}$ of non-local scaling operators $\phi_{\alpha}$ --scaling operators with a local core at the end of a semi-infinite line defect --, together with their scaling dimensions $\Delta_{\alpha}$,
\begin{equation} \label{eq:Rnl}
 \mathcal{R}' \ket{\phi_{\alpha}} = 2^{-\Delta_{\alpha}} \ket{\phi_{\alpha}}.
\end{equation}
For the critical Ising model, we find the scaling dimensions plotted on the third and fourth column in Fig. 5 of the main text, which includes accurate estimates of the scaling dimensions of the disorder $\mu$ and fermion $\psi$ and $\tilde{\psi}$ primary fields,
\begin{equation}
\Delta_\mu \approx 0.126,~~ \Delta_{\psi} \approx 0.49994, ~~\Delta_{\bar{\psi}} \approx 0.49994,
\end{equation}
very close to the exact solutions $0.125$, $0.5$, and $0.5$, respectively. 
 
\renewcommand\thefigure{C.\arabic{figure}}    
\renewcommand{\theequation}{C.\arabic{equation}}
\setcounter{figure}{0}  
\setcounter{equation}{0} 

\section{Section C: Fusion of local scaling operators and OPE coefficients.}

\begin{figure}[!t]
\begin{center}
\includegraphics[width=8cm]{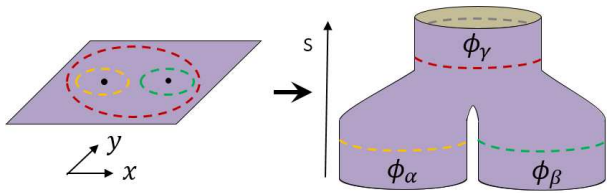}
\caption{
Schematics of a local scale transformation that maps the plane with two punctures into a pair of pants. We use this map to study the fusion of two primary fields $\phi_{\alpha}$ and $\phi_{\beta}$ into a third primary field $\phi_{\gamma}$, a process that occurs with an amplitude equal to the OPE coefficient $C_{\alpha \beta\gamma}$ \cite{CFT}.
} 
\label{fig:fusion}
\end{center}
\end{figure}
 	
In this section we explain how to compute the OPE coefficients of the CFT \cite{CFT} underlying a critical model. This will be accomplished by fusing together, through coarse-graining, the lattice representation $\ket{\phi_{\alpha}}$ and $\ket{\phi_{\beta}}$ of two primary operators $\phi_{\alpha}$ and $\phi_{\beta}$, and by decomposing the result in terms of the lattice representation $\ket{\phi_{\gamma}}$ of other scaling operators $\phi_{\gamma}$. For simplicity, we will assume that we already know which vectors $\ket{\phi_{\alpha}}$ correspond to primary fields of the theory. Our goal here is only to check that we can recover accurate estimates of the OPE for primary fields. [An approach that does not assume any previous knowledge of the primary fields of the underlying CFT will be presented elsewhere.] First we need to explain how to properly normalize the lattice representation $\ket{\phi_{\alpha}}$ of a primary operator $\phi_{\alpha}$. Then we will compute the OPE coefficients.

Recall that $\ket{\phi_{\alpha}}$ is obtained as a right eigenvector of the scaling map $\mathcal{R}$ that implements a translation on the cylinder or, equivalently, a change of scale on the plane, see Eq. 5 of the main text. The eigenvalue decomposition of $\mathcal{R}$ reads
\begin{equation} \label{eq:Rpi}
 \mathcal{R} = \sum_{\alpha} 2^{-\Delta_{\alpha}} \ket{\phi_{\alpha}}\bra{\pi_{\alpha}},
\end{equation}
where the kets $\{\ket{\phi_{\alpha}}\}$ and bras $\{\bra{\pi_{\alpha}}\}$ are the right and left eigenvectors of the linear map $\mathcal{R}$. These two sets of vectors form bi-orthonormal basis, that is $\braket{\pi_{\alpha}}{\phi_{\beta}} = \delta_{\alpha\beta}$. We recall also that Eq. 5 is obtained after normalizing the original tensors $A$ in the initial tensor network representation of $Z$, such that the largest eigenvalue of $\mathcal{R}$ is $2^{-\Delta_{\mathbb{I}}} = 2^{0} = 1$. In the following, we assume this normalization.

\begin{figure}[!t]
\begin{center}
\includegraphics[width=8cm]{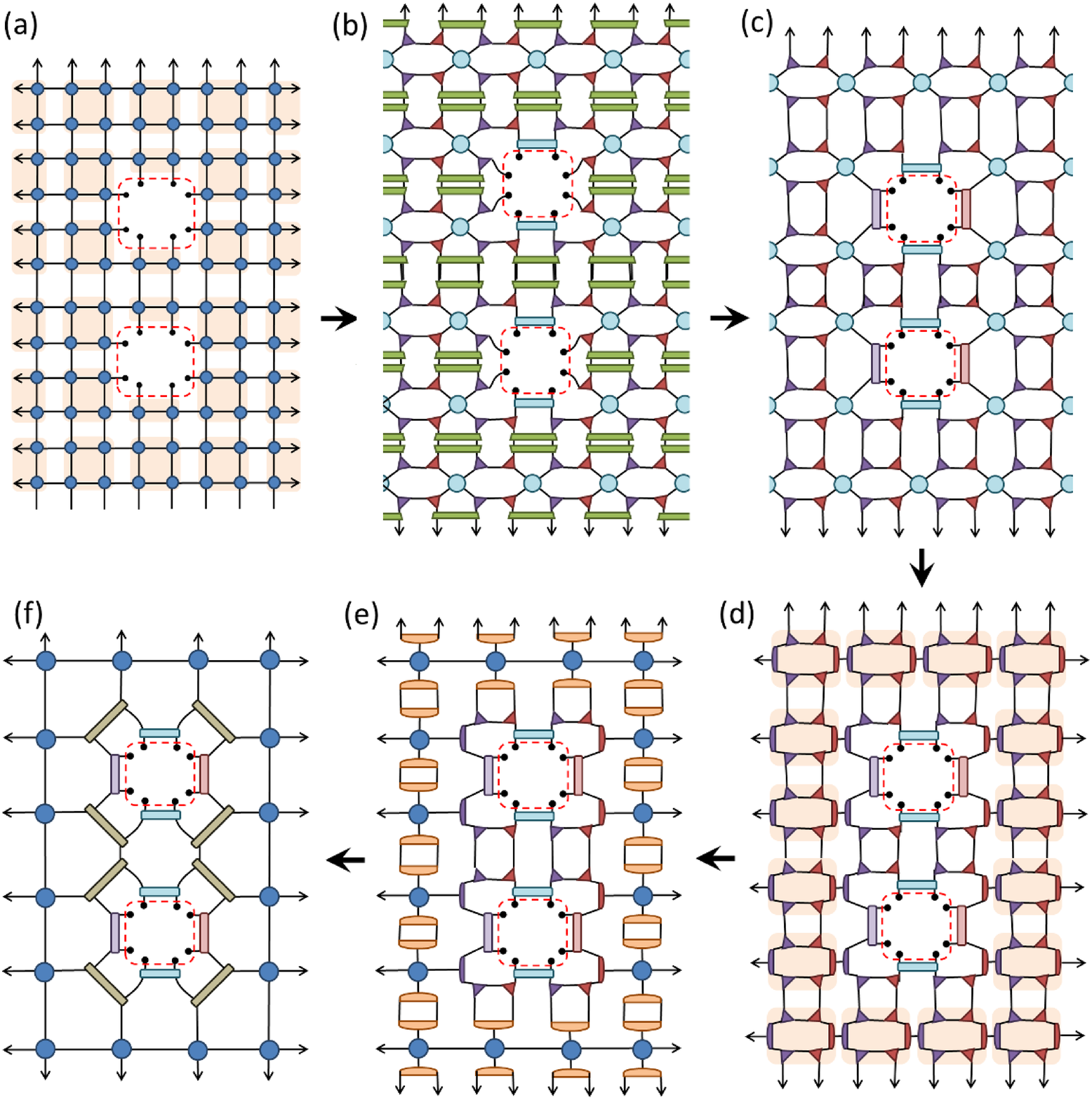}
\caption{ (a) Tensor network of the partition function $Z$ on the plane, with two punctures at $x=0$ and $y = \pm |y|$. 
(b)-(f) Detailed sequence of tensor replacements that bring two punctures closer (in this case, after the coarse-graining transformation the two punctures touch each other).
}
\label{fig:fusion0}
\end{center}
\end{figure}

It can be checked that the eigenvector of the scaling map $\mathcal{R}$ with largest eigenvalue $2^0=1$ corresponds to a block of $2\times 2$ of tensors $A$ (or coarse-grained versions of $A$). This eigenvector is the lattice representation $\ket{\mathbb{I}}$ of the identity primary field $\mathbb{I}$, since it corresponds indeed to inserting no operator (that is, the identity operator) in the partition function. This provides us with the natural normalization of $\ket{\mathbb{I}}$ and, through $\braket{\pi_{\mathbb{I}}}{\mathbb{I}}=1$, also of $\bra{\pi_{\mathbb{I}}}$.

Fig. \ref{fig:fusion} shows schematically the result of re-scaling a plane with two punctures, by an amount that is roughly proportional to the distance to the closest puncture. This results in a pair of pants, where each leg corresponds to one of the punctures. On the lattice, such transformation is obtained in three stages. 

The first stage is represented in Fig. \ref{fig:fusion0} and corresponds to distant punctures. In the current TNR scheme, it is computationally convenient to place the two punctures at the same value of $x$ (say $x=0$) and different value of $y = \pm |y|$ in the plane. We coarse-grain the partition function with two punctures applying the same rules as when we only had a single puncture at the origin: uniform coarse-graining of tensors $A$ everywhere in the plane except near the punctures, where the two sets of eight open indices are left untouched. The net result is as expected: a coarse-grained square tensor network of tensors $A'$, together with a double round of concentric tensors around each of the punctures, see Fig. \ref{fig:fusion0}(f). In addition, the two punctures are now effectively closer (meaning that $y' \approx y/2$). This is repeated until the two punctures have become nearest neighbours, which require about $\log_2 (y)$ iterations. 

\begin{figure}[!t]
\begin{center}
\includegraphics[width=8cm]{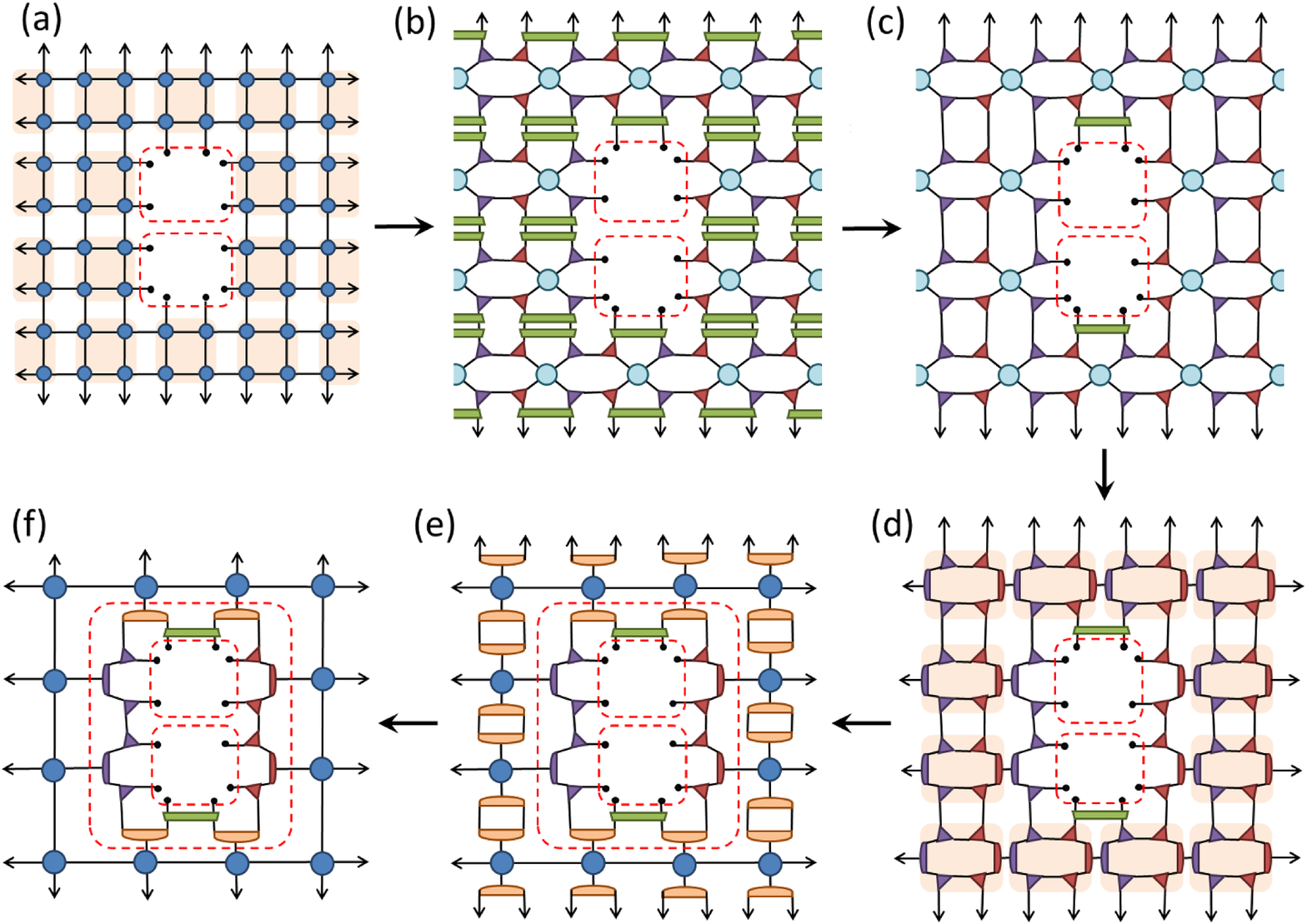}
\caption{
(a) Tensor network of the partition function $Z$ on the plane, with two adjacent punctures at $x=0$ and $y=\pm 1$ (see Fig. \ref{fig:fusion0}(f)). 
(b)-(f) Applying a homogeneous coarse-graining transformation everywhere in the plane except near the punctures, where the twelve open indices are left untouched, we obtain a coarse-grained tensor network with a single puncture (eight open indices) together with a linear map $\mathcal{F}$ that has fused the two punctures into one, see Fig. \ref{fig:fusion2}. 
}
\label{fig:fusion1}
\end{center}
\end{figure}

The second stage corresponds to the fusion of the two punctures into a single puncture. This is achieved by means of the coarse-graining transformation of Fig. \ref{fig:fusion1}. This transformation follows again the same rules as before (homogeneous coarse-graining everywhere except near the punctures, where the open indices are left untouched), and has the effect of indeed producing a single puncture with eight open indices at the origin, see Fig. \ref{fig:fusion1}(f). 

Finally, in the third stage, we continue to coarse-grain the plane with a single puncture, as we did in the main text. 

The tensor network resulting from these manipulations can be characterized as the product of two sets of linear maps $\{\mathcal{R}_{t}^{+}\}$ and $\{\mathcal{R}_{t}^{-}\}$ (one set for each puncture, but notice that $\mathcal{R}_{t}^{+} = \mathcal{R}_{t}^{-} = \mathcal{R}_{t}$) for the different scales $t$ involved in the first stage of coarse-graining, a fusion map $\mathcal{F}$ that transforms the product of two local operators into a single local operator for the second stage, and a set of linear maps $\{ \mathcal{R}_{t} \}$ for the scales $t$ of the third stage.

Notice that at criticality, scale invariance dictates that the linear maps $\{\mathcal{R}_{t}^{\pm}\}$ and $\{\mathcal{R}_{t}\}$ for the scales of stages one and three are the same scaling map $\mathcal{R}$ of Eq. \ref{eq:Rpi}. It is then useful to characterize the action of the fusion map $\mathcal{F}$ on the basis of right eigenvectors $\ket{\phi_{\alpha}}$ of $\mathcal{R}$, namely
\begin{equation} \label{eq:F}
\mathcal{F} \ket{\phi_{\alpha}} \otimes \ket{\phi_{\beta}} = \sum_{\gamma} D_{\alpha\beta}^{\gamma} \ket{\phi_{\gamma}}.
\end{equation}

We are finally ready to properly normalize the eigenvectors $\ket{\phi_{\alpha}}$ of the scaling map $\mathcal{R}$ corresponding to a primary field, say $\ket{\sigma}$, or $\ket{\epsilon}$, for the primary fields spin $\sigma$, and energy density $\epsilon$ of the Ising model. We do so by imposing that they fuse to the identity $\ket{\phi_{\mathbb{I}}}$ with unit amplitude, that is 
\begin{equation} \label{eq:normalizationOPE}
1 = D_{\alpha \alpha}^{\mathbb{I}} = \bra{\pi_{\mathbb{I}}} \mathcal{F} \ket{\phi_{\alpha}} \otimes \ket{\phi_\alpha}.
\end{equation}
Correspondingly, we also normalize $\bra{\pi_{\alpha}}$ so that $\braket{\pi_{\alpha}}{\phi_{\alpha}} = 1$.
Notice that the normalization of Eq. \ref{eq:normalizationOPE} is analogous to demanding that
\begin{equation}
\langle \phi_{\alpha}(z_1) \phi_{\alpha} (z_2) \rangle = \frac{1}{|z_1-z_2|^{2\Delta_{\alpha}}}
\end{equation}
in the continuum, as it is done in a CFT. For simplicity, we have assumed that the conformal spin $s_{\alpha}$ of $\phi_{\alpha}$ is zero. The more general case $s_{\alpha} \not = 0$ can also be dealt with by properly accounting for an additional phase $e^{-i\theta 2s_{\alpha}}$ in the correlator, where $(z_1-z_2)/|z_1-z_2| = e^{i\theta}$, and will be described elsewhere.

\begin{figure}[!t]
\begin{center}
\includegraphics[width=6cm]{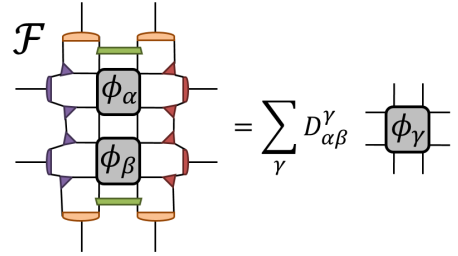}
\caption{
Tensor network representation of the fusion map $\mathcal{F}$ acting on the product $\ket{\phi_{\alpha}} \otimes \ket{\phi_{\beta}}$, see Eq. \ref{eq:F}.
}
\label{fig:fusion2}
\end{center}
\end{figure}

With the above normalization of $\ket{\phi_{\alpha}}$ and $\bra{\pi_{\alpha}}$ for each primary field $\phi_{\alpha}$, we can now obtain the operator product expansion (OPE) coefficients $C_{\alpha \beta \gamma}$ for primary fields $\phi_{\alpha}$, $\phi_{\beta}$, and $\phi_{\gamma}$ simply from
\begin{equation}
C_{\alpha\beta\gamma} = 2^{\Delta_{\gamma}}\bra{\pi_{\gamma}} \mathcal{F} \ket{\phi_{\alpha}} \otimes \ket{\phi_{\beta}}.
\end{equation}
In addition, we can add the non-local scaling operators (see Section B) to the mix. This requires also a generalized fusion map $\mathcal{F}'$ with suitable insertions of matrices implementing anti-periodic boundary conditions.

For the Ising model, using the tensors resulting from a simple $\chi=6$ TNR calculation, we obtain the following numerical estimates for the non-vanishing OPE coefficients involving both local (spin $\sigma$, energy density $\epsilon$) and non-local (disorder $\mu$, fermion $\psi$ and $\bar{\psi}$) primary fields,
\begin{eqnarray}
C_{\sigma\sigma\epsilon}\approx 0.5001,&& ~~ 
C_{\mu \mu \epsilon}\approx -0.4995, \\ 
C_{\mu\psi\sigma} \approx 0.9995 \frac{e^{-i\pi/4}}{\sqrt{2}},&&~~ 
C_{\mu\bar{\psi}\sigma} \approx 0.9995 \frac{e^{i\pi/4}}{\sqrt{2}}, \\
C_{\psi \bar{\psi} \epsilon} \approx -0.9998i,&& ~~ 
C_{\bar{\psi} \psi \epsilon  } \approx 0.9998i.
\end{eqnarray}
which are very close to the exact values 
\begin{eqnarray}
C_{\sigma\sigma\epsilon} = \frac{1}{2},&& ~~~~ 
C_{\mu \mu\epsilon } \approx -\frac{1}{2}, \\ 
C_{\mu\psi\sigma} = \frac{e^{-i\pi/4}}{\sqrt{2}},&&~~~~ 
C_{\mu\bar{\psi}\sigma} = \frac{e^{i\pi/4}}{\sqrt{2}}, \\
C_{\psi \bar{\psi} \epsilon} = -i,&& ~~~~ 
C_{\bar{\psi} \psi \epsilon} = i.
\end{eqnarray}


\renewcommand\thefigure{D.\arabic{figure}}    
\renewcommand{\theequation}{D.\arabic{equation}}
\setcounter{figure}{0}  
\setcounter{equation}{0} 

\section{Section D: Mapping of an n-sheeted Rieman surface into a cylinder.}

\begin{figure}[!t]
\begin{center}
\includegraphics[width=8cm]{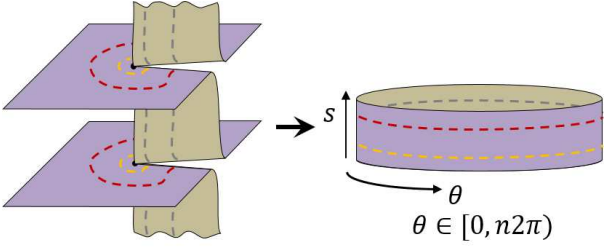}
\caption{
Conformal transformation that maps an $n$-sheeted Riemann surface, consisting of $n$ copies of the plane periodically connected through semi-infinite branch cuts along the positive $x$ axis (depicted is the $n=2$ case, with periodic boundary conditions on the vertical direction) to a cylinder ($s,\theta$) with $s\in \mathbb{R}$ and $\theta \in [0, n2\pi)$.
}
\label{fig:Riemann}
\end{center}
\end{figure}

\begin{figure}[!t]
\begin{center}
\includegraphics[width=8.5cm]{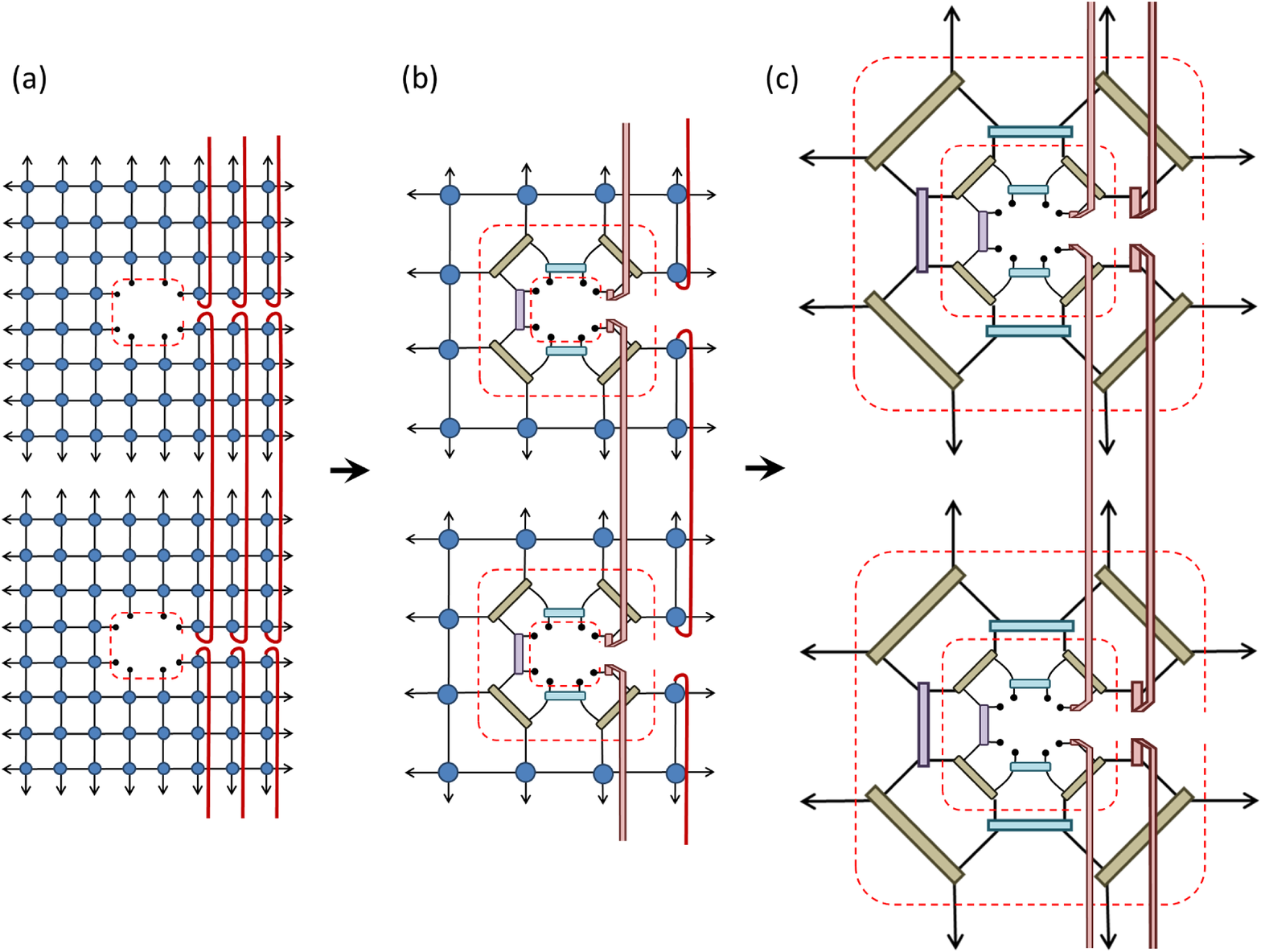}
\caption{
(a) Tensor network representation of the partition function $Z^{(n)}$ on $n$ copies of the plane connected into an $n$-sheeted partition function, and where we have added a puncture (removed four tensors) at the origin of each copy of the plane. Notice that the indices connecting tensors across the positive half of the $x$-axis have been cut and reconnected (red lines) with indices of tensors in other copies of the plane. We show the $n=2$ case, where periodic boundary conditions are assumed in the vertical direction, so that the open indices at the top are connected with the open indices at the bottom.
(b) The result of applying a homogeneous coarse-graining transformation throughout the tensor network except at the puncture is analogous to that in the plane, except that now it concerns $n$ connected copies and the double round of concentric tensors at the origin is made of $n$ times as many tensors.
(c) Iteration of the coarse-graining transformation produces a cylinder that can be understood as $n$ copies of the cylinder in the plane, properly cut and reconnected . That is, we can re-use all tensors obtained in coarse-graining the plane ($n=1$ case), so that no new numerical coarse-graining is actually needed in order to produce the radius-$n$ cylinder. 
}
\label{fig:Riemann1}
\end{center}
\end{figure}

In this section we consider a local scale transformation similar to the logarithmic conformal map discussed in the main text, but applied to an $n$-sheeted Riemann surface consisting of $n$ copies of the plane periodically connected through semi-infinite cuts along the positive $x$ axis, as used in the so-called replica method \cite{CFT,sReplica} to compute the trace of the $n^{th}$ power of the reduced density matrix of a CFT on a semi-infinite interval. The local scale transformation maps the $n$-sheeted Riemann surface to a cylinder of radius $n$. The case $n=2$ is schematically depicted in Fig. \ref{fig:Riemann1}, whereas the case $n=1$, which reduces to the plane, was already analysed in the main text.

Let $Z^{(n)}$ denote the partition function on this $n$-sheeted Riemann surface. A tensor network representation of $Z^{(n)}$ with a puncture at the origin is shown in Fig. \ref{fig:Riemann1}(a) for $n=2$. It is obtained from two copies of the tensor network for the punctured plane considered in the main text, see Fig. 2, by opening a cut on the positive $x$-axis of each copy of and reconnecting the open indices so as to suitably glue the different copies together.

\begin{figure}[!t]
\begin{center}
\includegraphics[width=8cm]{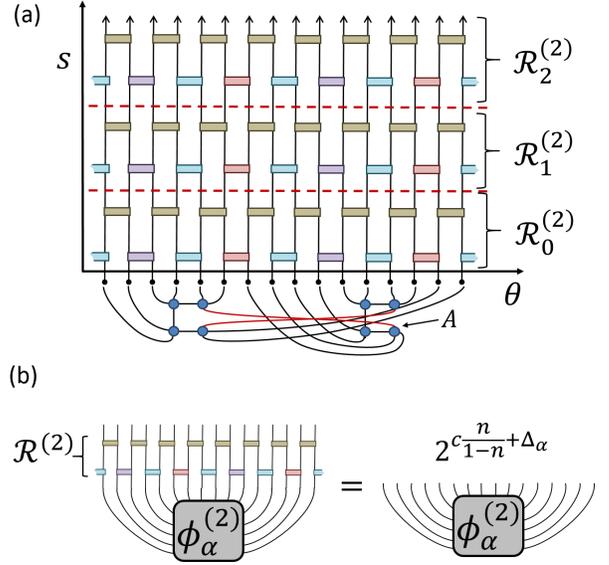}
\caption{
(a) Coarse-grained tensor network representation of the partition function $\mathcal{Z}^{(n)}$ (for $n=2$) in terms of a cylinder made of linear maps $\mathcal{R}^{(n)}_0$, $\mathcal{R}^{(n)}_1$, $\cdots$, and $n$ (duly interconnected) blocks of $2\times 2$ tensors $A$ at the bottom, filling the punctures introduced at the origin of each copy of the plane in Fig. \ref{fig:Riemann1}.
(b) Scaling map $\mathcal{R}^{(n)}$, whose eigenvectors are a lattice representation $\ket{\phi^{(n)}_{\alpha}}$ of the scaling operators $\phi^{(n)}_\alpha$ supported at the origin of the $n$ connected copies of the plane.  
}
\label{fig:Riemann2}
\end{center}
\end{figure}

We can now repeat the local scale transformation on the punctured plane, but adapted to the $n$-sheeted Riemann surface. Applying locally the same tensor replacements and manipulations that we used for the plane --but $n$ times, see Fig. \ref{fig:Riemann1}(b)-(c)--, we obtain a cylinder that is $n$ times as thick as when we started on the plane, see Fig. \ref{fig:Riemann2}.

The thick cylinder can be expressed as the product of linear maps $\mathcal{R}^{(n)}_0$, $\mathcal{R}^{(n)}_1$, $\cdots$, where the map $\mathcal{R}^{(n)}_{t}$ is made of the same tensors as the map $\mathcal{R}_{t}$ in the main text (but where each tensor appears $n$ times in $\mathcal{R}^{(n)}_{t}$  for each time that it appears in $\mathcal{R}_{t}$). The linear map $\mathcal{R}^{(n)}_{t}$ implements a change of scale on operators placed at the origin of the $n$ (connected) copies of the plane.
 
At criticality, the thick cylinder becomes translation invariant, in that $\mathcal{R}^{(n)}_{t}$ does not depend on the scale $t$. Let $\mathcal{R}^{(n)}$ denote the corresponding scaling map. Diagonilizing $\mathcal{R}^{(n)}$ we obtain a lattice representation $\ket{\phi_{\alpha}^{(n)}}$ of the scaling operators $\phi_{\alpha}^{(n)}$ leaving at the origin of the $n$ connected copies of the plane, 
\begin{equation} \label{eq:Rn}
 \mathcal{R}^{(n)} \ket{\phi_{\alpha}^{(n)}} = \lambda_{\alpha}^{(n)} \ket{\phi_{\alpha}^{(n)}},
\end{equation}
where now the eigenvalues 
\begin{equation}
\lambda_{\alpha^{(n)}} =  2^{-\frac{1}{n}\Delta_{\alpha} + \left(\frac{1}{n} -n\right) \frac{c}{12}}
\end{equation}
depend on the central charge $c$ of the CFT, the number of copies $n$, and the scaling dimension $\Delta_{\alpha}$ of a local scaling operator $\phi_{\alpha}$ in the plane (see Section E). 

When applied to the critical Ising model, from the eigenvalues $\lambda_{\alpha}^{(n)}$ for $n=2$ we were able to extract estimates of the scaling dimensions with the same accuracies as the ones from the plane discussed in the main text. In addition, we obtained a numerical estimate of the central charge $c\approx 0.5003$, which is close to the exact value $0.5$. One can additionally insert antiperiodic boundary conditions on one of the $n$ copies of the plane as discussed in Section B, to produce analogous results for the non-local scaling operators. 

We notice that the partition function $Z^{(n)}$ considered here is closely related to the computation of the entanglement Renyi entropy $S_n$ of order $n$ of a semi-infinite interval, 
\begin{equation}
S_n \equiv \frac{1}{1-n} \log \left( \tr (\rho^{n}) \right),
\end{equation}
where $\rho$ denotes the reduced density matrix for half of an infinite quantum spin chain, obtained from the ground state of an infinite quantum spin chain by tracing out half of the spins, and $\rho^n$ denotes its $n^{th}$ power. Specifically, we have that
\begin{equation}
S_n = \frac{1}{1-n} \log \left(  \frac{Z^{(n)}}{[Z^{(1)}]^n} \right),
\end{equation}
where $Z^{(1)}=Z$ is the partition function on the plane. It follows that the Renyi entropy for half an infinite interval can be expressed in terms of the dominant eigenvalue $\lambda^{(n)}_0$ of $\mathcal{R}^{(n)}$ as
\begin{equation}
S_n \approx \frac{1}{1-n} \log \left( \prod_{s=0}^{\infty} \lambda_0^{(n)}\right) =  
 \frac{1}{1-n} \sum_{s=0}^\infty \log \left( \lambda_0^{(n)} \right), 
\end{equation}
where we explicitly see that each scale $s$ contributes the same amount $\log (\lambda_0^{(n)})/(1-n)$ to $S_n$.


\renewcommand\thefigure{E.\arabic{figure}}    
\renewcommand{\theequation}{E.\arabic{equation}}
\setcounter{figure}{0}  
\setcounter{equation}{0} 

\section{Section E: CFT calculation of the transfer matrix spectrum}

In this section we consider the $z \rightarrow w = \log(z)$ conformal transformation that maps the plane into a cylinder in the continuum. By reproducing known results from conformal field theory, see e.g. Ref. \cite{CFT}, we first compute the exact spectrum of the transfer matrix on the resulting cylinder, which is given in terms of the scaling dimensions $\Delta_{\alpha}$ of the scaling operators $\phi_{\alpha}$ of the CFT. Then we also compute the exact spectrum of the transfer matrix on a thicker cylinder obtained from an $n$-sheeted Riemann surface by applying a similar conformal transformation. 

We emphasize that the CFT partition function only captures the universal part of the lattice partition function. As explained below, the non-universal, regular part of the partition function on the lattice (corresponding to the regular part of the free energy per site, denoted $a$ in Eq. \ref{eq:ZnonUni}) can be removed from the tensor network representation by changing the normalization of each tensor.

\subsection{From the plane to the cylinder, in the continuum.}

Recall that under a conformal transformation $z \rightarrow w = w(z)$, the (holomorphic part of the) stress tensor $T(z)$ of a CFT transforms as \cite{CFT}
\begin{equation} \label{eq:NewT}
 T'(w) = \left(\frac{dw}{dz}\right)^{-2} \left( T(z) -\frac{c}{12}\{w;z\} \right),
\end{equation}
where $\{w;z\}$ is the Schwarzian derivative
\begin{equation}
\{w;z\} = \frac{d^3w/dz^3}{dw/dz} - \frac{2}{3}\left( \frac{d^2w/dz^2}{dw/dz}\right)^2.
\end{equation}

Under the logarithmic transformation, mapping the plane into a cylinder, 
\begin{equation}
z \rightarrow w = \log (z),
\end{equation}
we have $dw/dz = 1/z$ and the Schwarzian derivative is $1/(2z^2)$. Therefore we find that the stress tensor $T_{cyl.}$ on the cylinder is related to the stress tensor $T_{pl.}$ on the plane by
\begin{equation}
T_{cyl.} (w) = z^2 T_{pl.}(z)  - \frac{c}{24}.
\end{equation}

Using the mode expansion \cite{CFT}
\begin{equation}
T_{pl.}(z) = \sum_{n\in \mathbb{Z}} L_n z^{-n-2},
\end{equation}
we find
\begin{equation}
T_{cyl.}(z) = \sum_{n\in \mathbb{Z}} L_n z^{-n} - \frac{c}{24}.
\end{equation}

The operators $L_0\pm\bar{L}_0$ generate dilations and rotations on the plane, and their spectra of eigenvalues correspond to the scaling dimensions $\Delta_{\alpha} = h_{\alpha} + \bar{h}_{\alpha}$ and conformal spins $s_{\alpha} = h_{\alpha} - \bar{h}_{\alpha}$ \cite{CFT}. Thus, in particular,
\begin{equation}
L_0+\bar{L}_0 = \sum_{\alpha} \Delta_{\alpha} \proj{\alpha}.
\end{equation}

On the cylinder, the generator of translations along the axis is written in terms of $L_0 + \bar{L}_0$ as
\begin{eqnarray}
H &\equiv& (L_0)_{cyl.} + (\bar{L}_0)_{cyl.} \\
&=& L_0 + \bar{L}_0 - \frac{c}{12} = \sum_{\alpha} \left(\Delta_{\alpha} -\frac{c}{12} \right)\proj{\alpha}.
\end{eqnarray}

\subsection{Back to the lattice.}

Let us now consider a critical (isotropic) lattice model on a torus of size $L_x \times L_y$, which corresponds to an imaginary modular parameter $\tau = i L_y/L_x$. The partition function $\tilde{Z}$ is expected to grow exponentially with the size of the torus according to
\begin{equation} \label{eq:ZnonUni}
\tilde{Z} \approx e^{a L_xL_y} Z(\tau),
\end{equation}
where $Z(\tau)$ is a universal contribution that depends on the modular parameter $\tau$, which for $\tau = i L_y/L_x$ reads \cite{CFT}
\begin{equation} \label{eq:Zuni}
Z(\tau) \equiv \tr \left( e^{-2\pi \frac{L_y}{L_x} H} \right) = \sum_{\alpha} e^{-2\pi \frac{L_y}{L_x} (\Delta_{\alpha} - \frac{c}{12})},
\end{equation}
and where the approximate ``$\approx$'' in Eq. \ref{eq:ZnonUni} signals that on the RHS we have neglected subleading, non-universal finite-size corrections. Notice that in the tensor network representation of the partition function $\tilde{Z}$ we can set the non-universal constant $a$ (which corresponds to the non-universal, regular part of the free energy) to zero by properly normalizing the tensors. For instance, if there is one tensor per site, then this is accomplished by dividing each tensor by $e^{-a}$. From now on, we assume that the tensors have been normalized in this way, so that
\begin{equation} \label{eq:Zuni2}
\tilde{Z} \approx Z(\tau),
\end{equation}
with the approximate ``$\approx$'' reminding us that $\tilde{Z}$ still contains non-universal, finite-size corrections to the universal part $Z(\tau)$ that become negligible for large $L_x$, $L_y$.

Notice that we can express the partition function $Z(\tau)$ as the trace of a power of a transfer matrix $\mathcal{M}$,
\begin{equation} 
Z(\tau) = \tr \left( \mathcal{M}^{L_y} \right),
\end{equation}
where the transfer matrix $\mathcal{M}$ implements translations by one lattice site in the $y$-direction,
\begin{equation} \label{eq:M2}
\mathcal{M} \equiv e^{-\frac{2\pi}{L_x} H} = \sum_{\alpha} e^{\frac{2\pi}{L_x} (\frac{c}{12} - \Delta_{\alpha})} \proj{\alpha}.
\end{equation}
We emphasize that the transfer matrix $\mathcal{M}$ is obtained by putting the original critical lattice model on a torus and taking a slice of its partition function, that is, without applying any local scale transformation. That the spectrum of the transfer matrix $\mathcal{M}$ is organized according to the scaling dimensions of the CFT is a well-know fact \cite{Cardy} which in the context of tensor network representations was first pointed out in Ref. \cite{TEFR} (see e.g. Refs. \cite{TEFR, TNR, sHauru} for a numerical confirmation). 

After transforming the discrete plane into a discrete cylinder by means of a lattice version of the conformal scale transformation $z \rightarrow w = \log (z)$, a translation on the resulting cylinder corresponds to a change of scale on the plane. Specifically, it corresponds to a change of scale by a factor $1/2$, and therefore we expect the transfer matrix $\mathcal{R}$ on this cylinder to have a spectral decomposition of the form
\begin{equation} \label{eq:Mcyl1}
\mathcal{R} = \sum_{\alpha} 2^{-\Delta_{\alpha}} \proj{\alpha}.
\end{equation}
On the other hand, the CFT calculation predicts that a translation on the cylinder corresponding to a change of scale on the plane by a factor $\lambda = 1/2$ is given by 
\begin{equation} \label{eq:Mcyl2}
e^{-2\pi \frac{\log(2)}{2\pi} H} = 2^{-H}  = \sum_{\alpha} 2^{ -\left( \Delta_{\alpha} - \frac{c}{12} \right)} \proj{\alpha},
\end{equation}
where the factor $\frac{\log(2)}{2\pi}$ can be understood as the ratio between the displacement $\log(2)$ on the cylinder, which corresponds to a change of scale on the plane, $e^{\log(2)} = 2$, and the radius $2\pi$ of the cylinder, which corresponds to a $2\pi$ rotation on the plane.

Eqs. \ref{eq:Mcyl1} and \ref{eq:Mcyl2} represent transfer matrices on the cylinder that only differ by a normalization constant, namely $2^{c/12}$. This normalization constant can be traced back to the contribution of the Schwarzian derivative in Eq. \ref{eq:NewT} \cite{sDavide}. That is, in the CFT, the transformation rule for the stress tensor includes a change in its normalization. On the lattice, the analogous change of normalization can be implemented by multiplying each tensor that appears in the transfer matrix by a suitable constant.

We thus conclude that the spectrum of eigenvalues of the scaling map $\mathcal{R}$ in Eq. \ref{eq:Mcyl1}, which is numerically obtained from TNR through a lattice version of the conformal transformation $z \rightarrow w = \log(w)$, corresponds to the spectrum of the transfer operator on the cylinder for a CFT, Eq. \ref{eq:Mcyl2}, up to an overall constant $2^{c/12}$.

\subsection{From the n-sheeted Riemann surface to the cylinder}

Let us now consider the $n$-sheeted Riemann surface described in Section D and Fig. \ref{fig:Riemann1}, which is obtained by suitably gluing together $n$ copies of the plane with a branching cut on the positive half of the real axis. By applying the logarithmic conformal transformation, we now obtain a cylinder with width $n2\pi$. This time, a translation on the cylinder by $\log(2)$ has a transfer matrix
\begin{eqnarray}
e^{-2\pi\frac{\log(2)}{n2\pi}H} =2^{-\frac{H}{n}} = \sum_{\alpha} 2^{-\frac{1}{n}\left( \Delta_{\alpha} - \frac{c}{12} \right)} \proj{\alpha}. \label{eq:Mncyl2}
\end{eqnarray}
where the factor $\frac{\log(2)}{n2\pi}$ can be understood again as the ratio between the displacement $\log(2)$ on the cylinder, which corresponds to a change of scale on the Riemann surface, $e^{\log(2)} = 2$, and the radius $n2\pi$ of the cylinder, which corresponds to a $n2\pi$ rotation on the Riemann surface.

On the other hand, the transfer matrix on the discrete cylinder obtained by TNR should have the spectrum of Eq. \ref{eq:Mncyl2} divided by a constant $(2^{c/12})^n$, that is,
\begin{equation} \label{eq:Mncyl1}
\mathcal{R}^{(n)} = \sum_{\alpha} 2^{-\frac{1}{n}\Delta_{\alpha} + \left(\frac{1}{n} -n\right) \frac{c}{12}} \proj{\alpha},
\end{equation}
which is indeed what we observed numerically, see Section D. We emphasize that this is the spectrum obtained by applying TNR without renormalizing the resulting tensors on the discrete cylinder. If we were now to renormalize those tensors to account for the Schwarzian derivative term in the transformation of the stress tensor, Eq. \ref{eq:NewT}, then the spectrum of Eq. \ref{eq:Mncyl2} would be obtained instead \cite{sDavide}. Notice that on the discrete cylinder of radius $n2\pi$ for $n>1$, the transfer matrix contains $8n$ tensors, corresponding to $n$ copies of the 8 tensors that appear in the transfer matrix for $n=1$. Then the change of normalization to account for the Schwarzian derivative can be implemented by multiplying each of these 8 tensors by $2^{(c/12)/8}$. Notice that this would consistently change the spectrum of the transfer matrix from Eq. \ref{eq:Mncyl1}
to Eq. \ref{eq:Mncyl2} for any $n\geq 1$.

\section{Section F: Use of a larger bond dimension}

In this section we discuss how the accuracy in the numerical estimates of the scaling dimensions $\Delta_{\alpha}$ (obtained by diagonalizing the scale transfer matrix $\mathcal{R}$) depends on the bond dimension $\chi$ used during the coarse-graining of the critical partition function $Z$ using tensor network renormalization.

Recall that the goal of this paper was to show that TNR can be used to implement a \textit{local} scale transformation on the lattice. In order to demonstrate that this is the case, we have considered a critical system, for which we have built a \textit{scale} transfer matrix $\mathcal{R}$ through a local scale transformation and shown that its spectrum of eigenvalues is closely related to the spectrum of eigenvalues of a \textit{space} transfer matrix $\mathcal{M}$. The relation between the eigenvalue spectra of $\mathcal{R}$ and $\mathcal{M}$ is a non-trivial consequence of local scale invariance of the critical theory in the continuum. We take the accurate realization of this relation on the critical lattice model as evidence that we have properly performed a local scale transformation on the lattice. 

From a more practical perspective, we can regard the connection between the eigenvalue spectra of the scale transfer matrix $\mathcal{R}$ and the space transfer matrix $\mathcal{M}$ as providing two alternative strategies to compute numerical estimates of the scaling dimensions $\Delta_{\alpha}$. The extraction of such numerical estimates from the eigenvalue spectrum of the space transfer matrix $\mathcal{M}$ has been extensively discussed in the literature \cite{TEFR,TNR,sHauru}. It was originally proposed in Ref. \cite{TEFR} for clean systems and was extended in Ref. \cite{sHauru} to systems with defects. On the other hand, the extraction of numerical estimates of the scaling dimensions $\Delta_{\alpha}$ from the eigenvalue spectrum of the scale transfer matrix $\mathcal{R}$ had not yet been explored, since it was not known how to produce $\mathcal{R}$ in the first place. In this paper we have explained how to build $\mathcal{R}$ and it is now natural to investigate how this second approach to extracting scaling dimensions compares to the first one.

\begin{table}[t]
\centering
\begin{tabular}{ccccc}
\; exact \; & \;\; TNR(6) \;\; & \; TNR(16) \; & \; TNR(24) \; & \; TRG(80) \; \\
\hline
0.125 & 0.125679 & 0.124941 & 0.124997 & 0.124989 \\
1     & 1.001499 & 1.000071 & 1.000009 & 1.000256 \\
1.125 & 1.125552 & 1.125011 & 1.124991 & 1.125532 \\
1.125 & 1.127024 & 1.125201 & 1.125027 & 1.125641 \\
2     & 2.003355 & 2.000087 & 2.000010 & 2.002235 \\
2     & 2.003365 & 2.000133 & 2.000017 & 2.002367 \\
2     & 2.003374 & 2.000279 & 2.000022 & 2.002607 \\
2     & 2.003525 & 2.000319 & 2.000060 & 2.003926 \\
2.125 & 2.114545 & 2.124944 & 2.124985 & 2.127266 \\
2.125 & 2.129043 & 2.125290 & 2.125038 & 2.130337 \\
2.125 & 2.142611 & 2.125670 & 2.125096 & 2.131299 \\
3     & 3.005045 & 3.000524 & 3.000052 & 3.007488 \\
3     & 3.005092 & 3.000777 & 3.000061 & 3.017253 \\
3     & 3.005259 & 3.000887 & 3.000073 & 3.017316 \\
3     & 3.005318 & 3.001010 & 3.000105 & 3.020581 \\
3     & 3.005805 & 3.001261 & 3.000206 & 3.023023 \\
3.125 & 3.109661 & 3.124866 & 3.124889 & 3.132764 \\
3.125 & 3.116466 & 3.125201 & 3.125019 & 3.132890 \\
3.125 & 3.118175 & 3.125319 & 3.125059 & 3.136725 \\
3.125 & 3.144798 & 3.126159 & 3.125099 & 3.137217 \\
3.125 & 3.145661 & 3.126163 & 3.125158 & 3.141363 \\
3.125 & 3.146323 & 3.126315 & 3.125172 & 3.146419 \\
\hline
\textrm{max err.} & 0.83\% & 0.046\% & 0.0069\% & 0.76\% \\      
\end{tabular}
\caption{Scaling dimensions $\Delta_\alpha$ of local scaling operators obtained from diagonalization of the transfer matrix $\mathcal R$ (on the cylinder that results from the logarithmic transform), which has been computed using TNR with bond dimensions $\chi = 6, 16, 24$. For comparison, the last column shows the spectrum of scaling dimensions obtained by diagonalizing the transfer matrix $\mathcal M$ on a finite system corresponding to $N=256$ spins, see Eq.8, using TRG with bond dimension $\chi = 80$. The bottom row denotes the maximum relative error in each of the columns.}
\label{tab:scdim}
\end{table}

For this purpose, we have built the scale transfer matrix $\mathcal{R}$ for the critical 2D classical Ising model for different values of the bond dimension $\chi=6,16,24$. [As explained in the main text, we first use TNR to apply a global scale transformation on the plane several times, realizing scale invariance. Then we build the scale transfer matrix $\mathcal{R}$ from the auxiliary tensors depicted in the inset of Fig. 2.] Tab. \ref{tab:scdim} shows the first 22 entries of the resulting spectrum of scaling dimensions $\Delta_{\alpha}$, which correspond to scaling dimension $\Delta \le 3.125$. The calculation with bond dimension $\chi=6$ required less than a minute to run on a laptop computer, as mentioned in the main text, while the calculation with bond dimension $\chi=24$ required approximately 2 hours to run on a laptop computer. For comparison, we also include scaling dimensions obtained by diagonalizing the space transfer matrix $\mathcal M$ on a finite system with linear size $N=256$ spins, using TRG with bond dimension $\chi=80$, which also required approximately 2 hours of computation time on the same laptop. 

As expected, the accuracy of the scaling dimensions extracted from $\mathcal R$ improves as we increase the bond dimension $\chi$. In the $\chi=6$ calculation the smallest 22 scaling dimensions $\Delta_\alpha$ are reproduced within $0.83\%$ accuracy, while in the $\chi=24$ calculation the error has been reduced by more than two orders of magnitude: the scaling dimensions $\Delta_\alpha$ are reproduced within an astounding $0.0069\%$ of accuracy. By comparison, the $\chi=80$ TRG calculation, which required comparable computation time to the $\chi=24$ TNR calculation, achieved only $0.76\%$ accuracy on the smallest 22 scaling dimensions. 

Alternatively, one may want to compare  the accuracy obtained through TNR calculations of the eigenvalue spectra of $\mathcal{R}$ (as shown here) and of $\mathcal{M}$ (see Appendix C of Ref. \cite{TNR}). We observe that for the same bond dimension, slightly better precision is still obtained when extracting the scaling dimensions from $\mathcal{R}$ than from $\mathcal{M}$. 

We conclude that the new approach to numerically extracting the scaling dimensions $\Delta_{\alpha}$, based on diagonalizing the scale transfer matrix $\mathcal{R}$, appears to produce more accurate estimates for the same computational cost, at least for the simple model under consideration. We emphasize, however, that from a computational viewpoint, the main advantage of the approach discussed in this paper is not that it yields improved estimates of the scaling dimensions $\Delta_{\alpha}$, but that it can be used to also extract the operator product expansion (OPE) coefficients of the underlying CFT (as discussed in some detail in Sect. C). Notice that we do not know how to compute the OPE coefficients from the space transfer matrix $\mathcal{M}$.


\begin{thebibliography}{99}
\bibitem{fpTEFR} When applied to the two-dimensional Ising model, tensor network renormalization \cite{TNR} was shown to explicitly recover scale invariance at the expected fixed points of the RG flow, including at the critical point. At non-critical fixed-points, TNR recovered the fixed-point tensors previously obtained by Gu and Wen's tensor entanglement-filtering renormalization (TEFR) \cite{TEFR}.
\bibitem{R} The linear map $\mathcal{R}_t$ is numerically seen to become approximately (but accurately) independent of $t$ only after suitably normalizing the original tensors $A$ (so as to eliminate the non-universal, extensive part of the free energy $\log (Z)$, see \SM \cite{SM} Section E, which would otherwise appear as a $t$-dependent multiplicative factor in $\mathcal{R}_t$), and fixing the gauge freedom in the tensor network, as explained in \cite{TNRalgorithms}.
\bibitem{M} The transfer matrix $\mathcal{M}$ provides an alternative, more direct way of computing the scaling dimensions $\Delta_{\alpha}$, Eq. \ref{eq:M} than transfer matrix $\mathcal{R}$, Eq. \ref{eq:R}, as already exploited \cite{TEFR, TNR, sHauru} (although the spectrum of $\mathcal{M}$ may be contaminated by non-universal finite-size corrections, not present in the spectrum of $\mathcal{R}$). However, from the eigenvectors $\ket{\phi_{\alpha}}$ of $\mathcal{R}$ we can compute the OPE coefficients.
\end{thebibliography}
\end{document}